\documentclass[12pt]{article}

\usepackage{amssymb}

\begin{document}

%%%%%%%%%%%%%%%%%%%%%%%%%%%%%%%%%%%%%%%%%%%%%%%%%%%%
\newcommand{\PV}{{\rm PV}}
\newcommand{\E}{\mathbb{E}}
\newcommand{\T}{\mathbb{T}}
\newcommand{\R}{\mathbb{R}}
\newcommand{\Z}{\mathbb{Z}}
\newcommand{\N}{\mathbb{N}}

\newcommand{\beqn}{\begin{eqnarray}}
\newcommand{\eeqn}{\end{eqnarray}}
\newcommand{\be}{\begin{equation}}
\newcommand{\ee}{\end{equation}}
\newcommand{\ba}{\begin{array}}
\newcommand{\ea}{\end{array}}

\newcommand{\ds}{\displaystyle}
\newcommand{\pa}{\partial}
\newcommand{\ve}{\varepsilon}
\newcommand{\supp}{\mathop{\rm supp}\nolimits}

\newtheorem{theorem}{Theorem}[section]
\renewcommand{\theequation}{\thesection.\arabic{equation}}
\newtheorem{definition}[theorem]{Definition}
\newtheorem{lemma}[theorem]{Lemma}
\newtheorem{remark}[theorem]{Remark}
\newtheorem{cor}[theorem]{Corollary}
\newtheorem{pro}[theorem]{Proposition}

\newcommand{\bo}{{\hfill\loota}}
\newcommand{\loota}{\hbox{\enspace{\vrule height 7pt depth 0pt width 7pt}}}
%%%%%%%%%%%%%%%%%%%%%%%%%%%%%%%%%%%%%

\begin{titlepage}
\hspace{2cm}
 \begin{center}
{\Large\bf On convergence to equilibrium for one-dimensional chain
 of harmonic oscillators on the half-line }
\vspace{1cm}\\
{\large T.V. Dudnikova}\medskip\\
{\it  Keldysh Institute of Applied Mathematics,\\
Miuskaya sq. 4, Moscow 125047, Russia}\medskip\\
E-mail:~tdudnikov@mail.ru
\end{center}
\vspace{1cm}

\begin{abstract}
The initial-boundary value problem for an infinite
one-dimensional chain of harmonic oscillators
 on the half-line is considered.
The large time asymptotic behavior of solutions is studied.
The initial data of the system are supposed to be a
random function which has  some mixing properties.
  We study the distribution $\mu_t$
  of the random solution at time moments $t\in\R$.
  The main result is the convergence of $\mu_t$
   to a Gaussian probability measure as $t\to\infty$.
We find stationary states in which there is a non-zero energy current at origin.
\medskip

\end{abstract}
\end{titlepage}

%%%%%%%%%%%%%%%%%%%%%%%%%%%%%%%%%%%%%
\section{Introduction}
%%%%%%%%%%%%%%%%%%%%%%

We consider the infinite system of harmonic oscillators
on the half--line:
\beqn
\ddot u(x,t)=(\Delta_L-m^2) u(x,t),\quad x\in\N,\quad t>0,
\label{1.1}
\eeqn
with the boundary condition (as $x=0$)
\beqn\label{1.2}
\ddot u(0,t)=F(u(0,t))-m^2u(0,t)-\gamma\dot u(0,t)+u(1,t)-u(0,t),
\quad t>0,
\eeqn
and with the initial condition (as $t=0$)
\beqn\label{1.3}
u(x,0)=u_0(x),\quad \dot u(x,0)=v_0(x),\quad x\in\Z_+.
\eeqn
Here $u(x,t)\in\R$, $m\ge0$,  $\gamma\ge0$, $\Z_+=\{0,1,2,3,\dots\}=0\cup\N$,
 $\Delta_L$ denotes the second derivative on $\Z$:
$$
\Delta_L u(x)=u(x+1)-2u(x)+u(x-1),\quad x\in\Z.
$$
If $\gamma=0$, then formally  system (\ref{1.1}), (\ref{1.2}) is
Hamiltonian with  the Hamiltonian functional
 \beqn\label{H}
 {\rm H}(u,\dot u):= \frac{1}{2}
\sum\limits_{x\in\Z_+}\Big(
|\dot u(x,t)|^2+|u(x+1,t)-u(x,t)|^2+m^2|u(x,t)|^2\Big)+P(u(0,t)),
\eeqn
where, by definition,
$P(q):=-\int F(q)\,dq$, $q\in\R$.
To prove the existence of solutions to the problem (\ref{1.1})--(\ref{1.3}),
we assume that $P\equiv0$ or
\be\label{P}
P\in C^2(\R),\quad
P(q)\to+\infty\quad \mbox{as }\,|q|\to\infty,
\ee
so $P(q)\ge P_0$ for all $q$ with some $P_0\in\R$.
%%%%%%%%%%%%%%%%%%%
\medskip

Write $Y(t)=(Y^0(t),Y^1(t))\equiv(u(\cdot,t),\dot u(\cdot,t))$,
$Y_0=(Y_0^0,Y_0^1)\equiv(u_0(\cdot),v_0(\cdot))$.
We assume that the initial state $Y_0(x)$ belongs to
the Hilbert space ${\cal H}_{\alpha,+}$, $\alpha\in\R$, consisting of
real sequences, see Definition \ref{d1.1'} below.
The existence and uniqueness of the solutions $Y(t)$
is proved in Appendix~A.

To prove the main result
we assume that $F(q)=-\kappa q$ with $\kappa\ge0$.
Moreover, we impose some restrictions on the coefficients $m,\kappa,\gamma$
of the system, see condition {\bf C} below. In particular,
if $\gamma\not=0$, then at least one of the numbers $m$ and  $\kappa$ is not zero.
If $\gamma=0$, then $\kappa\in(0,2)$.
The initial state $Y_0(x)$ is supposed to be a
random element of the space ${\cal H}_{\alpha,+}$, $\alpha<-3/2$,
with the distribution $\mu_0$.
We assume that $\mu_0$ is a probability measure
of mean zero satisfying conditions {\bf S2}--{\bf S4}. In
particular,
 the initial measure $\mu_0$ satisfies a mixing condition.
Roughly speaking, it means that
$Y_0(x)$ and $Y_0(y)$ are asymptotically  independent as $|x-y|\to\infty$.

For a given $t\in\R$, denote by $\mu_t$ the probability measure on
${\cal H}_{\alpha,+}$ giving the distribution of the random
solution $Y(t)=(u(\cdot,t),\dot u(\cdot,t))$ to the problem (\ref{1.1})--(\ref{1.3}).
Our main objective is  to prove the weak convergence of the measures $\mu_t$
on the space ${\cal H}_{\alpha,+}$ with $\alpha<-3/2$
to a limit measure $\mu_{\infty}$, which is
 an equilibrium Gaussian measure on ${\cal H}_{\alpha,+}$,
\begin{equation}\label{1.8i}
  \mu_t \rightharpoondown \mu_\infty\quad  \mbox{as }\,\,t\to \infty.
  \end{equation}
  This means the convergence of the integrals
  $\displaystyle\int f(Y)\,\mu_t(dY) \to \int f(Y)\,\mu_\infty(dY)$
  as $t\to\infty$ for any bounded continuous functional
  $f$ on ${\cal H}_{\alpha,+}$.
Furthermore, we find stationary states $\mu_\infty$
in which there is a non-zero energy current at the origin,
see Remark~\ref{remark2.11} below.

 For  one-dimensional chains of harmonic
oscillators in the whole line, the convergence to equilibrium distribution
 has  been established by Boldrighini {\it et al.} \cite{BPT}
 and by Spohn and Lebowitz  \cite{SL}.
Ergodic properties of one-dimensional chains of anharmonic
oscillators coupled to heat baths were studied by Jak\v{s}i\'c, Pillet
and others (see, e.g., \cite{JP97, JP98, EPR}).
 The convergence (\ref{1.8i}) was proved also
  for harmonic crystals in $\Z^d$ with $d\ge1$ \cite{DKS1, DKM} and
 for a scalar Klein--Gordon field coupled to a harmonic crystal \cite{DK}.
 Similar results were obtained in \cite{D08} for the harmonic crystals
 in the half-space $\Z^d_+:=\{x=(x_1,\dots,x_d)\in\Z^d:\,x_1\ge0\}$
with zero boundary condition.
In the present paper, we treat the one-dimensional model
but with the boundary condition of the form (\ref{1.2}).
\medskip

We outline the strategy of the proof.
Using the technique of~\cite{DKS1,D08}, we derive
 the convergence (\ref{1.8i}) from  the assertions {\bf I} and {\bf II}:\\
{\bf I.} The family of measures $\mu_t$, $t\geq 0$, is weakly
compact in ${\cal H}_{\alpha,+}$ with $\alpha<-3/2$.\\
{\bf II.}
The characteristic functionals of $\mu_t$ converge to a Gaussian functional,
\be\label{2.6i}
 \hat\mu_t(\Psi):= \int \exp({i\langle Y,\Psi\rangle_+})\,\mu_t(dY)
\rightarrow \displaystyle \exp\{-\frac{1}{2}{\cal Q}_\infty (\Psi,\Psi)\}, \quad t\to\infty.
\ee
Here $\Psi=(\Psi^0,\Psi^1) \in{\cal S}:=S\oplus S$,
where $S$ denotes a space of real rapidly decreasing sequences,
$$
\langle Y,\Psi \rangle_+
=\sum\limits_{i=0,1}\sum\limits_{x\in\Z_+} Y^i(x)\Psi^i(x)\quad
\mbox{for }\,\,Y=(Y^0,Y^1)\in{\cal H}_{\alpha,+}\,\,\,\mbox{ and }\,\,
 \Psi=(\Psi^0,\Psi^1) \in{\cal S},
$$
${\cal Q}_\infty(\Psi,\Psi)$ is a quadratic form.
\smallskip

To prove {\bf I} we derive the uniform bound (\ref{c.1})
for the mean local energy with respect to the measure
 $\mu_t$, $t\ge0$, and apply the Prokhorov compactness theorem.
To check {\bf II} we study the asymptotic behavior of the solution
$Y(t)$ and obtain (see Lemma~\ref{l6.7}) that
\be\label{0.1}
\langle Y(t),\Psi\rangle_+\sim \langle U_0(t)Y_0,\Pi_\Psi\rangle,
\quad t\to\infty\quad \mbox{(in mean)},
\ee
where $U_0(t)$ is a solving operator to the problem
 (\ref{1.1}) and (\ref{1.3}) with zero boundary condition (i.e., $u(0,t)\equiv0$),
the functions $\Pi_\Psi$ are expressed using $\Psi\in{\cal S}$
(see formula (\ref{hn})).
\smallskip

To prove (\ref{0.1})
we decompose the solution $u(x,t)$ into two terms:
$u(x,t)=z(x,t)+q(x,t)$.
Here $z(x,t)$ is a solution of (\ref{1.1}) satisfying zero boundary condition
and the initial condition (\ref{1.3}), i.e.,
$(z(\cdot,t),\dot z(\cdot,t))=U_0(t)Y_0$.
$q(x,t)$ is a solution of (\ref{1.1}) with zero initial condition for $x\not=0$
and
with the boundary condition (\ref{b.3}).
The existence and behavior of the solution $z(x,t)$ were studied
in \cite{D08}.
We state the basic results on $z(x,t)$  in Section~\ref{sec2.3}.
For $q(x,t)$ with $x\not=0$, the following representation holds,
\be\label{1.7}
q(x,t)=\int_{0}^t K(x,t-s)q(0,s)\,ds,\quad x\in\N,
\ee
where the kernel $K(x,t)$ is defined in (\ref{K(x,t)}).
The properties of this kernel are studied in Appendix~A.
In particular, $K(x,t)$ satisfies the following bound:
\be\label{0.2}
\sum\limits_{x\in\N}(1+x^2)^{\alpha} |K(x,t)|^2
\sim C(1+|t|)^{-3} \quad \mbox{as }\,\,t\to\infty,
\quad \mbox{for any }\,\,\alpha<-3/2.
\ee
Therefore, to prove (\ref{0.1}) it remains to study
the long time behavior
of $q(x,t)$ with $x=0$. By (\ref{1.2}) and (\ref{1.7}),
$q(0,t)$ evolves according to a Volterra integro-differential equation
of the form
\be\label{int7.1}
\ddot q(0,t)=F(q(0,t))-(1+m^2)q(0,t)-\gamma \dot q(0,t)+\int_0^t
K(1,t-s)q(0,s)\,ds+h(t),\quad t>0,
\ee
with $h(t)=z(1,t)$.
To investigate the solutions of (\ref{int7.1}),
we assume that $F(q)=-\kappa q$ with $\kappa\ge0$.
This allows us to apply the Fourier--Laplace transform $t\to\omega$
to the solutions $q(0,t)$
and study the analytic properties of $\tilde q(0,\omega)$
for complex values $\omega\in\mathbb{C}$ (see Appendix~B).
These properties give the bound for the solutions $q(0,t)$
of (\ref{int7.1}) with $h(t)\equiv0$:
$$
|q(0,t)|\le C(1+t)^{-3/2}, \quad t\ge0.
$$
Applying this estimate together with (\ref{1.7}) and (\ref{0.2}),
we obtain the asymptotics for $q(x,t)$ in mean (see formula (\ref{6.15}) below)
$$
\langle q(\cdot,t),\psi\rangle_+\sim\langle U_0(t)Y_0,{\bf K}^0_\psi\rangle_+,
 \quad t\to\infty,
$$
where $\psi\in S$, the vector valued function ${\bf K}^0_\psi$ is defined
in (\ref{Kpsi}).
This implies the asymptotics (\ref{0.1}) which plays the crucial role
in our convergence analysis
for the statistical solutions of the problem (\ref{1.1})--(\ref{1.3}).
\medskip

The  dynamics of the equations
with delay has been extensively investigated by many authors
 under some restrictions on the kernel $K(1,t)$.
 For details, we refer to the monograph by Gripenberg, Londen and Staffans  \cite{GLS}.
The stability properties of linear Volterra integro-differential equations
can be found in the papers by Murakami \cite{Murakami},
Nino and Murakami \cite{NinoMur}, % Hara \cite{Hara},
 Kordonis and Philos \cite{KorPhi}.
Note that in the literature frequently
the long time asymptotical behavior of solutions
%%of linear Volterra integro-differential equations
 is studied assuming that the kernel  has the exponential decay or
is of one sign. However, in our case, by (\ref{0.2}),
 the decay of $K(1,t)$ is like $(1+t)^{-3/2}$.
\medskip

The nonlinear equations of the form (\ref{int7.1})
  with a stationary Gaussian process $h(t)$ and with a smooth
 (confining or periodic) potential $P(q)=-\int F(q)\,dq$
  have been investigated  extensively, for example,
   the ergodic properties of such equations were studied
 by Jak\v{s}i\'c and  Pillet in \cite{JP97},
 and the qualitative properties of solutions were established by Ottobre and Pavliotis in \cite{OP}.
In the present paper, we prove the convergence to equilibrium
for a linear model. However,
we do not assume that the initial distribution of the system is a
Gibbs measure or absolutely continuous with respect to a Gibbs
measure. Therefore,
the force $h(t)=z(1,t)$ in (\ref{int7.1})
is non-Gaussian, in general.
\medskip
%%------------------------------------------

 The paper is organized as follows.
 In Section~\ref{sec2}, we impose the conditions on the model and on
the initial measures $\mu_0$ and state the main results.
The bounds for solutions of (\ref{int7.1})
are obtained in Section~\ref{sec3}.
 The asymptotics (\ref{0.1}) is proved in Section~\ref{sec5}.
 In Section~\ref{s5}, we  prove the convergence (\ref{1.8i}).
In Appendix~A, we study the properties of the kernel  $K(x,t)$
and prove the existence of the solutions.
The properties of $\tilde q(0,\omega)$, $\omega\in\mathbb{C}$, are studied in Appendix~B.
 Appendix~C contains the results on the solutions $z(x,t)$.

%%%%%%%%%%%%%%%%%%%%%%%%% Section 2   %%%%%%%%%%%%%%%%%%%%%%%%%%%%%%%%%
\setcounter{equation}{0}
\section{Main Results}\label{sec2}
%%-------------------------------------------
\subsection{Phase space}
%%------------------------

 We assume that the initial date $Y_0$
belongs to the phase space ${\cal H}_{\alpha,+}$,
 $\alpha\in\R$, defined below.
%%%%%%%%%%%%%%%%%%%%%%%%%%%%%%%%%%%%%%
 \begin{definition} \label{d1.1'}
(i) $\ell^2_{\alpha,+}\equiv\ell^2_{\alpha,+}(\Z_+)$,
$\alpha\in\R$, is the  Hilbert space of sequences $u(x)$, $x\in\Z_+$,
 with the norm
$$ \Vert u\Vert^2_{\alpha,+}
=\sum\limits_{x\in\Z_+}|u(x)|^2\langle x\rangle^{2\alpha}<\infty,\quad
\langle x\rangle:=(1+x^2)^{1/2}.
$$
(ii)
$ {\cal H}_{\alpha,+}=\ell^2_{\alpha,+}\otimes\ell^2_{\alpha,+}$
is the  Hilbert space of pairs $Y(x)=(u(x),v(x))$ of sequences equipped with the norm
 $ \Vert Y\Vert^2_{\alpha,+}
= \Vert u\Vert^2_{\alpha,+}+ \Vert v\Vert^2_{\alpha,+}<\infty$.
\smallskip\\
(iii) $\ell^2_{\alpha}\equiv\ell^2_{\alpha}(\Z)$
 is the  Hilbert space of sequences with the norm
$\Vert u\Vert^2_{\alpha}
=\sum\limits_{x\in\Z}|u(x)|^2 \langle  x\rangle^{2\alpha}<\infty$. In particular,
$\ell^2_0\equiv \ell^2$.
Write ${\cal H}_{\alpha}:=\ell^2_{\alpha}\otimes\ell^2_{\alpha}$, $\alpha\in\R$.
 \end{definition}
%%%%%%%%%%%%%%%%%%%%%%%%%%%%%%%%%%%%%
\begin{theorem}\label{T.A}
Let $\gamma,m\ge0$ and condition (\ref{P}) hold,
 and let $Y_0\in{\cal H}_{\alpha,+}$, $\alpha\in\R$.
Then the problem (\ref{1.1})--(\ref{1.3}) has
a unique solution $Y(t)\in C(\R,{\cal H}_{\alpha,+})$. The operator
$U(t):Y_0\to Y(t)$ is continuous on ${\cal H}_{\alpha,+}$. Moreover,
 there exist constants $C,B<\infty$
such that $\Vert U(t)Y_0\Vert_{\alpha,+}\le Ce^{B|t|}$,
where the constant $C$ depends on $\Vert Y_0\Vert_{\alpha,+}$.
\end{theorem}
%%%%%%%%%%%%%%%%%%%%%%%%%%%%%%%%%%

Theorem~\ref{T.A} is proved in Appendix~A.
The proof is based on the following representation for
 the solution $u(x,t)$ of the problem (\ref{1.1})--(\ref{1.3}):
\be\label{2.1}
u(x,t)=z(x,t)+q(x,t),\quad x\in\Z_+,\quad t>0,
\ee
where $z(x,t)$  is a solution of the initial-boundary value problem
with zero boundary condition,
\beqn
&&\ddot z(x,t)=(\Delta_L-m^2)z(x,t),\quad x\in\N,\quad t>0,
\label{a.1}\\
&&z(0,t)=0,\quad t\ge0,
\label{a.2}\\
&&z(x,0)=u_0(x),\quad \dot z(x,0)=v_0(x),\quad x\in\N.
\label{a.3}
\eeqn
%%-----------------
 Therefore, $q(x,t)$  is a solution of the following initial-boundary value problem
\beqn
&&\ddot q(x,t)=(\Delta_L-m^2)q(x,t),\quad x\in\N,\quad t>0,\label{b.1}\\
&&\ddot q(0,t)=F(q(0,t))-m^2 q(0,t)-\gamma\dot q(0,t) +q(1,t)-q(0,t)
+z(1,t),\,\, t>0, \label{b.2}\\
&&q(x,0)=0,\quad \dot q(x,0)=0,\quad x\in\N, \label{b.3}\\
&&q(0,0)=u_0(0),\quad \dot q(0,0)=v_0(0). \label{b.4}
\eeqn
%%------------------

In Section \ref{sec2.3}, we state the properties of the solutions $z(x,t)$.
In Section \ref{sec3},  we study the behavior of  $q(x,t)$.

To prove the main result we assume that $F(q)=-\kappa q$, with $\kappa\ge0$.
Moreover, on the nonnegative constants $m,\gamma,\kappa$ of the system we impose
the following condition {\bf C}.
\begin{description}
\item[${\bf C}$]
 If $\gamma\not=0$, then either $m\not=0$ or  $\kappa\not=0$.
In addition, if  $\gamma\in(0,1)$ and $m=0$, then $\kappa\not=2(1-\gamma^2)$;
 if $\gamma\in\left(0,\left(\sqrt{m^2+4}-m\right)/2\right]$ and $m\not=0$,
 then $\kappa\not=1-\gamma^2\pm\sqrt{(1-\gamma^2)^2-m^2\gamma^2}$.
 If $\gamma=0$, then $\kappa\in(0,2)$.
\end{description}

%%%%%%%%%%%%%%%%%%%%%%%%%%%    2.2     %%%%%%%%%%%%%%%%%%%%%%%%%%%%%%%
\subsection{Random initial data}\label{sec2.2}
%%%%%%%%%%%%%%%%%%%%%%%%%%%%%%%%%%%%%%%%%%%%%%%%%%

We assume that the initial date $Y_0(x)=(Y_0^0(x),Y_0^1(x))$
is a measurable random function with values in
$({\cal H}_{\alpha,+},{\cal B}({\cal H}_{\alpha,+}))$, where
${\cal B}( {\cal H}_{\alpha,+})$ stands for the Borel $\sigma$-algebra in ${\cal H}_{\alpha,+}$.
Denote by $\mu_0$ a Borel probability measure on ${\cal H}_{\alpha,+}$
 giving the distribution of $Y_0$.
The expectation with respect to $\mu_0$ is denoted by $\E$.
We impose the following conditions {\bf S1}--{\bf S4} on $\mu_0$.
\begin{description}
\item{\bf S1}
The measure $\mu_0$ has zero expectation value,
$\ds\E(Y_0(x))\equiv\int Y_0(x)\mu_0(dY_0)=0$, $x\in\Z_+$.

\item{\bf S2}
 The measure $\mu_0$  has finite variance,
$$\sup\limits_{x\in\Z_+}\E(|Y_0(x)|^2)\le e_0<\infty.$$
\end{description}

%%%%%%%%%%%%%%
Write $\nu_0=\mu_0\{Y_0\in{\cal H}_{\alpha,+}:\, Y_0(0)=0\}$.
The expectation with respect to $\nu_0$ is denoted by $\E_0$.
The correlation functions of the measure $\nu_0$ are denoted by
$$
Q^{ij}_0(x,x')=\E_0(Y^i_0(x)Y^j_0(x')),\quad x,x'\in\Z_+,
\quad i,j=0,1.
$$
In particular, the matrix $Q_0(x,x')=(Q^{ij}_0(x,x'))_{i,j=0,1}$
vanishes if $x=0$ or $x'=0$.
On the measure $\nu_0$ we impose conditions {\bf S3} and {\bf S4}.
%%-----------------------------------------------------------
\begin{description}
\item{\bf S3} For every $x\in\Z$,
$\lim\limits_{y\to +\infty} Q^{ij}_0(x+y,y)=q^{ij}_0(x)$, $i,j=0,1$.
Here $q^{ij}_0(x)$ are correlation functions of some translation
invariant measure $\nu$ with zero mean value on the space ${\cal H}_{\alpha}$.
\end{description}
%%-------------------------------
By definition, a measure $\nu$ is said to be {\em translation invariant} if $\nu(T_hB)=\nu(B)$
for any $B\in{\cal B}({\cal H}_\alpha)$ and $h\in\Z$, where
$T_h Y_0(x)=Y_0(x-h)$ for $x\in\Z$.
\smallskip

To formulate the last condition on $\nu_0$,
denote by ${\cal A}$ an open interval of $\Z_+$ and by $\sigma ({\cal A})$
the $\sigma $-algebra on ${\cal H}_{\alpha,+}$ generated
 by $Y_0(x)$ with $x\in{\cal A}$.
Define the Ibragimov mixing coefficient
of a probability  measure  $\nu$ on ${\cal H}_{\alpha,+}$
by the rule (cf \cite[Definition 17.2.2]{IL})
$$
\varphi(r)\equiv
\sup_{\small{\ba{c} {\cal A},{\cal B}\subset \Z_+\\
{\rm dist} ({\cal A},{\cal B})\ge r\ea}}
\sup_{\small{
\ba{c} A\in\sigma({\cal A}),B\in\sigma({\cal B})\\ \nu(B)>0\ea}}
\frac{| \nu(A\cap B) - \nu(A)\nu(B)|}{ \nu(B)}.
$$
 A measure $\nu$ is said to satisfy the  {\em strong uniform
Ibragimov mixing condition} if $\varphi(r)\to 0$ as $r\to\infty$.
\begin{description}
\item{\bf S4} The measure $\nu_0$ satisfies the strong
uniform Ibragimov mixing condition with mixing coefficient $\varphi$,
and
$\displaystyle\int_{0}^{+\infty}\varphi^{1/2}(r)\,dr<\infty$.
\end{description}

The examples of the initial measures  $\nu_0$ satisfying conditions {\bf S3} and {\bf S4}
are given in \cite{D08}.
%%%%%%%%%%%%%%%%%%%%%%%%%%%%%%%%%%%%%%%%%%%%%%%%%%%
\begin{lemma}\label{lcom} (see \cite[Proposition 5.2]{D08})
Let conditions {\bf S1}--{\bf S4} hold.
Then  $q_0^{ij}\in\ell^1$. Moreover,
for any $\Phi,\Psi\in {\cal H}_{0,+}$,
 the following bound holds
\be\label{2.42}
|\langle Q_0(z,z'),\Phi(z)\otimes\Psi(z')\rangle_+|\le
C\Vert\Phi\Vert_{0,+} \Vert\Psi\Vert_{0,+}.
\ee
\end{lemma}
%%%%%%%%%%%%%%%%%%%%%%%%%%%%%%%%%%%%%%%%%%%%%%%%%%%%%%%%%%%%%%%%

This implies, in particular, that
 $\hat q^{ij}_0\in C(\T)$.
 Here and below $\hat q(\theta)\equiv F_{x\to\theta} [q(x)]$ denotes the
 discrete Fourier transform w.r.t. $x\in\Z$,
 $$
 \hat q(\theta)=\sum\limits_{x\in\Z} e^{ix\theta} q(x),
 \quad \theta\in\T\equiv \R/2\pi\Z.
 $$

For a  probability  measure $\nu$ on  ${\cal H}_{\alpha,+}$
 denote by $\hat\nu$ the characteristic functional (Fourier transform)
$$
\hat\nu(\Psi)\equiv\int\exp(i\langle Y,\Psi\rangle_+ )\,\nu(dY),\quad
 \Psi\in {\cal S}.
$$
A  measure $\nu$ is called {\em Gaussian} (with zero expectation) if
its characteristic functional has the form
$\hat {\nu} (\Psi)=\exp\{-{\cal Q}(\Psi,\Psi)/2\}$, $\Psi \in {\cal S}$,
where ${\cal Q}$ is a  real nonnegative quadratic form in ${\cal S}$.

%%%%%%%%%%%%%%%%%%%%%%%%%%   2.3   %%%%%%%%%%%%%%%%%%%%%%%%%%%
\subsection{Convergence to equilibrium for the problem (\ref{a.1})--(\ref{a.3})}
\label{sec2.3}
%%%%%%%%%%%%%%%%%%%%%%%%%%%%%%%%%%%%%%%%%%%%%%%%%%%%%%%%%%%%%%

In this section we state the results concerning the solutions of
the problem (\ref{a.1})--(\ref{a.3}) (see \cite{D08}).
Write $Z(t)\equiv Z(x,t)=(z(x,t),\dot z(x,t))$.
%%%%%%--------------------------------------------------
\begin{lemma} \label{l2.1} (see Lemma 2.7 in \cite{D08})
Assume that $\alpha\in\R$. Then
(i) for any  $Y_0 \in {\cal H}_{\alpha,+}$, there exists  a unique solution
$Z(t)\in C(\R, {\cal H}_{\alpha,+})$  to the  problem
(\ref{a.1})--(\ref{a.3});
(ii) the operator  $U_0(t):Y_0\mapsto Z(t)$ is continuous
on ${\cal H}_{\alpha,+}$.
\end{lemma}
%%-------------------

The proof is based on the following formula for
the solution $Z(x,t)$ of the problem (\ref{a.1})--(\ref{a.3}):
\be\label{sol}
Z^i(x,t)=\sum\limits_{j=0,1}\sum\limits_{x'\in\N}
{\cal G}^{ij}_{t,+}(x,x') Y_0^j(x'), \quad x\in \Z_+,
\ee
where $Z(x,t)=(Z^0(x,t),Z^1(x,t))\equiv(z(x,t),\dot z(x,t))$,
the Green function ${\cal G}_{t,+}(x,x')$ is
\beqn\label{3.2}
 {\cal G}_{t,+}(x,x'):={\cal G}_t(x-x')-{\cal G}_t(x+x'),
\quad
{\cal G}_t(x)\equiv\frac1{2\pi}\int\limits_{\T}
e^{-ix \theta} \hat{\cal G}_t(\theta)\,d\theta,\quad x\in\Z,
\eeqn
with
\beqn\label{hatcalG}
\hat {\cal G}_t(\theta)=(\hat{\cal G}_t^{ij}(\theta))_{i,j=0}^1
=\left(\ba{ll} \cos\phi(\theta)t&
\frac{\sin\phi(\theta)t}{\phi(\theta)}\\
-\phi(\theta)\sin\phi(\theta)t& \cos\phi(\theta)t
\ea\right),
\,\,\, \phi(\theta)=\sqrt{2-2\cos\theta+m^2}.
\eeqn
In particular,
$\phi(\theta)=2|\sin(\theta/2)|$ if $m=0$.
We see that $Z(0,t)\equiv 0$ $\forall t$, since
${\cal G}_t(-x)={\cal G}_t(x)$.
%%%%%%%%%%%%%%%%%%%%%%%%%%%%%%%%
\begin{definition}\label{def2.6}
We define $\nu_t$ as the Borel probability measure on
${\cal H}_{\alpha,+}$, which gives the distribution of the solution $Z(t)$
to the problem (\ref{a.1})--(\ref{a.3}), i.e.,
$\nu_t(B)=\nu_0(U_0(-t)B)$ for any $B\in{\cal B}({\cal H}_{\alpha,+})$, $t\in\R$.
\end{definition}
%%%%--------------------------
\begin{lemma}\label{l1} (see \cite[Theorem A]{D08})
The correlation functions of $\nu_t$ converge to a limit,
\be\label{concor}
 Q^{ij}_t(x,x')=\int Z^i(x) Z^j(x')\,\nu_t(dZ)\to
 Q^{ij}_\infty(x,x'), \quad t\to\infty,\quad x,x'\in\Z_+.
\ee
The correlation matrix $Q_\infty(x,x')=(Q^{ij}_\infty(x,x'))_{i,j=0}^1$
has the form
\be\label{correlation}
Q_\infty(x,x')=q_\infty(x-x')-q_\infty(x+x')-q_\infty(-x-x')+
q_\infty(-x+x'),\quad x,x'\in\Z_+.
\ee
Here $q_\infty(x)=q^+_\infty(x)+q^-_\infty(x)$, $x\in\Z$,
where in the Fourier transform we have
\beqn\label{q-infty}
\ba{rcl}
\hat q^+_\infty(\theta)&=&\frac12\left(\ba{cc}
\hat q_0^{00}(\theta)+\hat q_0^{11}(\theta)\phi^{-2}(\theta)&
\hat q_0^{01}(\theta)-\hat q_0^{10}(\theta)\\
-\hat q_0^{01}(\theta)+\hat q_0^{10}(\theta)&\phi^2(\theta)\hat q_0^{00}(\theta)+
\hat q_0^{11}(\theta)\ea\right),\\~\\
\hat q^-_\infty(\theta)&=&
\frac1{2}\,{\rm sign}(\theta)\left(\ba{cc}
\left(\hat q_0^{10}(\theta)-\hat q_0^{01}(\theta)\right)\phi^{-1}(\theta)&
\hat q_0^{00}(\theta)\phi(\theta)+\hat q_0^{11}(\theta)\phi^{-1}(\theta)\\
-\hat q_0^{00}(\theta)\phi(\theta)-\hat q_0^{11}(\theta)\phi^{-1}(\theta)
&(\hat q_0^{10}(\theta)-\hat q_0^{01}(\theta))\phi(\theta)\ea\right).
\ea
\eeqn
Here $\theta\in\T$ if $m\not=0$ and $\theta\in\T\setminus\{0\}$ if $m=0$,
   the functions $q_0^{ij}$,
   $i,j=0,1$,   are introduced in condition {\bf S3}, $\phi(\theta)$
   is defined in (\ref{hatcalG}).
\end{lemma}
%%%%%%%%5----------------------------------------------------------------

Denote by
${\cal Q}^\nu_{\infty} (\Psi,\Psi)$ a real quadratic form on
 ${\cal S}$  defined by
\beqn\label{qpp}
{\cal Q}^\nu_{\infty} (\Psi,\Psi)=
\langle Q_{\infty}(x,x'),\Psi(x)\otimes \Psi(x')\rangle_+\equiv
\sum\limits_{i,j=0,1}\sum\limits_{x,x'\in\Z_+}
Q^{ij}_{\infty}(x,x')\Psi^i(x) \Psi^j(x').
\eeqn
%%-----------------------------------------------
\begin{remark}\label{rem2.8}
{\rm
Given $\Psi(x)$, $x\in\Z_+$,
introduce  an odd sequence $\Psi_{o}(x)$, $x\in\Z$,
such that $\Psi_o(x)=\Psi(x)$ for
$x>0$, $\Psi_o(0)=0$ and $\Psi_o(x)=-\Psi(-x)$ for $x<0$.
Note that $Q_{\infty}(x,x')$ is odd w.r.t. $x$ and $x'$.
Moreover,
\be\label{2.24}
{\cal Q}^\nu_{\infty} (\Psi,\Psi)
=\langle q_\infty(x-x'),\Psi_o(x)\otimes\Psi_o(x')\rangle
\equiv\sum\limits_{i,j=0}^{1}\sum\limits_{x,x'\in\Z}
q_\infty^{ij}(x-x')\Psi_o^{i}(x)\Psi_o^{j}(x').
\ee
In the case $m\not=0$, $q^{ij}_\infty\in\ell^1$,
by Lemma \ref{lcom} and formulas (\ref{q-infty}).
Then, Young's inequality yields
$$
|{\cal Q}^\nu_{\infty} (\Psi,\Psi)|\le C\Vert q_\infty\Vert_{\ell^1}
\Vert\Psi_o\Vert^2_{0}\le C_1\Vert q_\infty\Vert_{\ell^1}
\Vert\Psi\Vert^2_{0,+},
$$
i.e., ${\cal Q}^\nu_{\infty} (\Psi,\Psi)$ is continuous in
${\cal H}_{0,+}\equiv\ell^2_{0,+}\times\ell^2_{0,+}$.
Here as before $\Vert\cdot\Vert_0$ ($\Vert\cdot\Vert_{0,+}$)
stands for the norm in $\ell^2_0\equiv\ell^2$ and in ${\cal H}_{0}$
(in $\ell^2_{0,+}$ and in ${\cal H}_{0,+}$, respectively).

In the case $m=0$, $\phi(\theta)=2|\sin(\theta/2)|$.
Note that
$|\hat\Psi_o(\theta)|\le C(\Psi)|\sin\theta|$ for $\Psi\in S$.
Denote
$\ell^2_{1/\phi,+}=\{\Psi\in\ell^2_{0,+}: \hat\Psi_o/\phi\in L^2(\T)\}$
with  norm
$\Vert \Psi\Vert_{1/\phi,+}:=\Vert(1+1/\phi(\theta))\hat\Psi_o(\theta)\Vert_{L^2(\T)}$.
Therefore, for any
$\Psi=(\Psi^0,\Psi^1)\in{\cal H}_{1/\phi,+}:=\ell^2_{1/\phi,+}\times\ell^2_{0,+}$,
$$
{\cal Q}^\nu_{\infty} (\Psi,\Psi)\le \Vert q_0\Vert_{\ell^1}
\left(C_1\Vert F^{-1}(1/\phi)*\Psi_o^0\Vert_0^2+C_2\Vert\Psi_o\Vert^2_{0}\right)
\le C_3\Vert\Psi^0\Vert^2_{1/\phi,+}+C_4\Vert\Psi^1\Vert^2_{0,+},
$$
by (\ref{q-infty}) and (\ref{2.24}).
Here and below by $F^{-1}$ we denote the inverse Fourier transform.
}\end{remark}
%%%-----------------------------------

\begin{theorem}\label{t1}
Let conditions {\bf S1}--{\bf S4} hold and let
$\alpha<-1/2$ if $m\not=0$ and $\alpha<-1$ if $m=0$.
 Then the following assertions are fulfilled.
 \begin{description}
\item{(i)} The measures $\nu_t$ weakly converge
  as $t\to\infty$ on the space ${\cal H}_{\alpha,+}$.
   The limit measure $\nu_\infty$ is Gaussian, with zero mean value
 and with correlation matrix $Q_\infty$ defined in (\ref{correlation}).

 \item{(ii)} The uniform bound holds,
\beqn \label{ubound}
\sup\limits_{t\ge 0}
\E_0\Vert U_0(t)Y_0\Vert^2_{\alpha,_+}<\infty.
\eeqn
Moreover,
$\lim_{t\to\infty}\E_0|\langle U_0(t)Y_0,\Psi\rangle_+|^2
={\cal Q}^\nu_\infty(\Psi,\Psi)$ for any $\Psi\in {\cal S}$.

\item{(iii)} The limit measure $\nu_\infty$ satisfies a mixing condition
w.r.t. the group $U_0(t)$, i.e.,
for any $f,g\in L^2({\cal H}_{\alpha,+},\nu_\infty)$,
$\displaystyle\lim_{t\to\infty}\int f(U_0(t)Y)g(Y)\nu_\infty(dY)
=\int f(Y)\nu_\infty(dY)\int g(Y)\nu_\infty(dY)$.
\end{description}
\end{theorem}
%%%%%%%%%%%%%%%%%%%%%%%%%

If $m\not=0$, then the results of
Lemmas \ref{l1} and Theorem \ref{t1} follow directly from \cite{D08}.
In the case $m=0$,
the proof of the bound (\ref{ubound}) needs in some modification
(see Appendix~C).

%%%%%%%%%%%%%%%%%%%%%%%%%%   2.4   %%%%%%%%%%%%%%%%%%%%%%%%%%%
\subsection{Main theorem}
%%%%%%%%%%%%%%%%%%%%%%%%%%%%%%%%%%%%%%%%%%%%%%%%%%%%%%%%%%%%%%

Denote by ${\cal Q}_{\infty} (\Psi,\Psi)$
a real quadratic form in ${\cal S}$ of the form
\beqn\label{Qmu}
{\cal Q}_{\infty} (\Psi,\Psi)=
{\cal Q}^\nu_{\infty} (\Pi_\Psi,\Pi_\Psi),
\eeqn
where ${\cal Q}_\infty$  is defined by (\ref{qpp}), $\Pi_\Psi$ is introduced in (\ref{hn}).
%%%-----------------------------------------
\begin{remark}\label{rem2.11}
For any $\Psi\in{\cal S}$, $\Pi_\Psi\in {\cal H}_{1/\phi,+}$.
This follows from (\ref{hn}) and Remark~\ref{rem6.6}.
Therefore, by Remark \ref{rem2.8} and formula~(\ref{Qmu}),
${\cal Q}_\infty(\Psi,\Psi)$ is bounded for all $\Psi\in{\cal S}$.
\end{remark}

 We  define $\mu_t$ as the Borel probability measure on ${\cal H}_{\alpha,+}$
given the distribution of the solution $Y(t)$ to the problem (\ref{1.1})--(\ref{1.3}),
$\mu_t(B) = \mu_0(U(-t)B)$, where $B\in {\cal B}({\cal H}_{\alpha,+})$
and $t\in \R$.
%-----------------------------------------------
The main result of the paper is the following theorem.
%%-------------------------------
\begin{theorem}\label{t0}
Let conditions {\bf S1}--{\bf S4}  hold
and the constants $m,\gamma,\kappa$ satisfy condition {\bf C}. Then
the following assertions are fulfilled.
\begin{itemize}
\item[(i)] The measures $\mu_t$ weakly converge as $t\to\infty$
to a limit measure $\mu_\infty$
on the space ${\cal H}_{\alpha,+}$ with $\alpha<-3/2$.
Moreover, the limit measure is Gaussian  with zero mean value.

\item[(ii)] The correlation functions of $\mu_t$
converge to a limit, and for $\Psi\in{\cal S}$,
$$
\lim_{t\to\infty}\E|\langle Y(t),\Psi\rangle_+|^2
={\cal Q}_\infty(\Psi,\Psi).
$$
\end{itemize}
\end{theorem}

This theorem is proved in Section~\ref{s5}.
%%%%%%%%%%%%%%%%%%%%%%%%%%% add
\begin{remark}\label{remark2.11}
{\rm The limiting energy current at the origin equals
$J:=-\gamma \lim\limits_{t\to\infty}\E\left(\dot u(0,t)\right)^2$.
Hence, in the case when $\gamma>0$, $J\not=0$
if $\int(Y^1(0))^2\,\mu_\infty(dY)\not=0$.
The limit measures $\mu_\infty$ satisfying the last condition can be constructed as follows.

Let the initial correlations functions $q_0^{ij}$ from condition {\bf S3} have a form
$q_0^{ij}\equiv0$ for $i\not=j$, and $q_0^{00}\cdot q_0^{11}\not\equiv0$.
Write $r(\theta):=\phi^2(\theta)\hat q_0^{00}(\theta)+\hat q_0^{11}(\theta)$,
$\theta\in\T$. Then, $r\in C(\T)$ (see Lemma~\ref{lcom}), $r(\theta)\ge0$,
and $r(\theta)\not\equiv0$. Moreover,
$\hat q_\infty^{11}(\theta)=r(\theta)/4$, by (\ref{q-infty}).
Using (\ref{4.10'}), we obtain
$$
\int(Y^1(0))^2\,\mu_\infty(dY)=\lim_{t\to\infty}\E\left(\dot u(0,t)\right)^2=
\frac1{2\pi}\int_{\T}\sin^2(\theta)\,r(\theta)\,
\Big|\int_0^{+\infty}N(s)e^{i\phi(\theta)s}\,ds\Big|^2\,d\theta\not=0.
$$
The properties of $N(t)$ are studied in Appendix~B.
}\end{remark}

%%%%%%%%%%%%%%%%%%%%%%%   Section 3 %%%%%%%%%%%%%%%%%%%%
\setcounter{equation}{0}
\section{Bounds for $q(x,t)$} \label{sec3}
%%%%%%%%%%%%%%%%%%%%%%%%%%%%%%%%%%%%%

In this section, we investigate the behavior of the solutions  $q(x,t)$
to the problem (\ref{b.1})--(\ref{b.4})  as $t\to\infty$.
At first, we study the properties of $q(x,t)$ with $x\not=0$
using the Fourier--Laplace transform.
%%-------------------------------------------------------
\begin{definition}
Let $|q(t)|\le Ce^{B t}$.
The Fourier--Laplace transform of $q(t)$ is given by the formula
\be\label{La-F}
\tilde q(\omega)=\int_{0}^{+\infty}
e^{i\omega t}q(t)\,dt,\quad \Im\omega>B.
\ee
\end{definition}
%%----------------------------

The Gronwall inequality and conditions on $F(q)$
imply standard a priori estimates for the solutions $q(x,t)$, $x\in\Z_+$.
In particular,
there exist constants $A,B<\infty$ such that
$$
\sum\limits_{x\in\Z_+}(|q(x,t)|^2+|\dot q(x,t)|^2)\le C e^{Bt}
\quad\mbox{ as }\,\, t\to+\infty.
$$
This bound is proved in Appendix A (see formulas (\ref{7.9}) and (\ref{7.10})).
Hence the Fourier--Laplace  transform with respect to $t$-variable,
$q(x,t)\to\tilde q(x,\omega)$, exists at least
for $\Im \omega>B$ and satisfies the following equation
\beqn\label{b.1'}
(-\Delta_L+m^2-\omega^2)\tilde q(x,\omega)=0,
\quad x\in\N,\quad \Im\omega>B,
\eeqn
by (\ref{b.3}). Now  we construct the solution of (\ref{b.1'}).
We first note that the Fourier transform of the lattice operator $-\Delta_L+m^2$
is the operator of multiplication by the function
$\phi^2(\theta)=2-2\cos\theta+m^2$. Thus, $-\Delta_L+m^2$ is a self-adjoint operator
and its spectrum is absolutely continuous and coincides with the range of $\phi^2(\theta)$,
i.e., with $[m^2,m^2+4]$.
Second, the following lemma holds (see Lemma 2.1 in  \cite{KKK}).

%%%%%%%%%%%%%%%%%%%%%%%%%%%%%%%%%%%%%%%%%%%%%%%%%%%%
\begin{lemma}\label{theta}
Denote $\Lambda:=[-\sqrt{4+ m^2},-m]\cup[m,\sqrt{4+m^2}]$.
For given $\omega\in \mathbb{C}\setminus \Lambda$,
the equation
\be\label{32}
2-2\cos\theta+m^2=\omega^2
\ee
has the unique solution  $\theta(\omega)$ in the domain
$\{\theta\in\mathbb{C}:\,\,\Im\theta>0,\,\,
-\pi<\Re\theta\le\pi\}$.
Moreover, $\theta(\omega)$ is an analytic function in $\mathbb{C}\setminus \Lambda$.
\end{lemma}
%%%%%%%%%%%%%%%%%%%%%%%%%%%%%%%%%%%%%%%%%%%%%%%%%%%%%%%%%%%%%%%%%%%%%%%

Since we seek the solution  $q(\cdot,t)\in \ell^2_{\alpha,+}$
with some $\alpha$, $\tilde q(x,\omega)$ has a form
$$
\tilde q(x,\omega)=\tilde q(0,\omega) e^{i\theta(\omega)x},
\quad x\in\N.
$$
We put $\tilde K(x,\omega)=e^{i\theta(\omega)x}$.
Applying the inverse Fourier--Laplace transform with respect to $\omega$-variable,
we write the solution $q(x,t)$ of (\ref{b.1}) in the form
\be\label{qxt}
(q(x,t),\dot q(x,t))=
\int_{0}^t K(x,t-s)\big(q(0,s),\dot q(0,s)\big)\,ds,\quad x\in\N,\quad t>0,
\ee
where
\be\label{K(x,t)}
K(x,t)=\frac1{2\pi}\int\limits_{-\infty+i\mu}^{+\infty+i\mu}
e^{-i\omega t}\tilde K(x,\omega)\,d\omega,\quad \tilde K(x,\omega)=e^{i\theta(\omega)x},\quad x\in\N,\quad t>0,
\ee
%%%%%%%%%%%%%%%%%%%%%%%%%%%%%%%%%%%%%%%%%%%%%%%%%%%%%%%%%%%%%%%%%%%%%%
%%and $q(\cdot,t)\in\ell^2_{\alpha,+}$.
%Moreover, the left-hand side of (\ref{2.23}) does not depend on $\mu$.
%The integral in (\ref{2.23}) is understood in the sense of principal value.
%%To study the large time behavior of $K(x,t)$,
%% we will move down the contour of integration in  (\ref{K(x,t)}).
%%--------------------------------------------------------------------------------
with some $\mu>0$. In Appendix A, we study the analytic properties
$\tilde K(x,\omega)$ for $\omega\in\mathbb{C}$ and obtain the following result.
\begin{theorem}\label{l2.15}
For any $\alpha<-3/2$, the following bound holds,
 \be\label{boundK}
 \Vert K(\cdot,t)\Vert'_{\alpha,+}:=
 \sqrt{\sum\limits_{x\in\N}\langle x\rangle^{2\alpha}|K(x,t)|^2}\le C(1+t)^{-3/2}
 \quad for \quad t>0.
 \ee
In particular,
\be\label{3.7'}
|K(1,t)|\le C(1+t)^{-3/2},\quad t>0.
\ee
\end{theorem}
%%%---------------------------------------------------------------------------------------------

To estimate $q(0,t)$, we again apply the Fourier--Laplace transform.
Using (\ref{qxt}) and equality $F(q)=-\kappa q$ with $\kappa\ge0$, we
rewrite Eqn~(\ref{b.2}) in the form
\be\label{7.1}
\ddot q(0,t)=-(\kappa+1+m^2)q(0,t)-\gamma \dot q(0,t)+\int_0^t
K(1,t-s)q(0,s)\,ds+z(1,t).
\ee
At first, we study the solutions of
  the corresponding homogeneous equation
\be\label{qt}
\ddot q(0,t)=-(\kappa+1+m^2)q(0,t)-\gamma \dot q(0,t)+\int_0^t
K(1,t-s)q(0,s)\,ds,\quad t>0,
\ee
with the initial data
 \be\label{init}
q(0,t)|_{t=0}=u_0(0)=: q_0,\quad \dot q(0,t)|_{t=0}=v_0(0)=: p_0.
\ee
%%----------------------
Applying the Fourier--Laplace transform
to the solutions $q(0,t)$ of (\ref{qt}), we obtain
$$
\tilde D(\omega)\tilde q(0,\omega)
=-i\omega q_0+q_0\gamma +p_0 \quad \mbox{for }\,\,\Im\omega>B,
$$
where, by definition,
\begin{equation}\label{3.21}
\tilde D(\omega):=-\omega^2+\kappa+1+m^2-i\omega\gamma
-\tilde K(1,\omega),\quad
\tilde K(1,\omega)=e^{i\theta(\omega)}.
\end{equation}
Write $ \tilde  N(\omega) :=[\tilde D(\omega)]^{-1}$.
Then
$\tilde q(0,\omega)=\tilde N(\omega)\left(-i\omega q_0+q_0\gamma +p_0\right)$.
The analytic properties of $\tilde D(\omega)$ and $\tilde N(\omega)$
for $\omega\in\mathbb{C}$ are studied in Appendix B.
In particular, we prove that $\tilde N(\omega)$ is analytic in the upper half-space,
i.e., with $\Im\omega>0$.
Denote
\be\label{N}
N(t)=\frac1{2\pi}\int\limits_{-\infty+i\mu}
^{+\infty+i\mu}e^{-i\omega t} \tilde N(\omega)\,d\omega,\quad t\ge0,
\quad \mbox{with some }\,\mu>0.
\ee
The following theorem is proved in Appendix~B.
%%--------------------------------------------------
\begin{theorem}\label{l3.1}
Let constants $m,\gamma,\kappa$ be nonnegative and
satisfy condition {\bf C}. Then
\be\label{NN}
|N^{(k)}(t)|\le C(1+t)^{-3/2},\quad t\ge0,\quad k=0,1,2.
\ee
\end{theorem}
%------------------------------

If  condition {\bf C} is not satisfied, then
 $N(t)$ decreases more slowly than $\langle t\rangle^{-3/2}$,
see Remark~\ref{rem-b}.
We need the bound (\ref{NN}) to derive asymptotics (\ref{5.4})
which plays the crucial role in our proof.
%%--------------------------
\begin{cor} \label{cor3}
Denote by $S(t)$ a solving operator of the Cauchy problem
(\ref{qt}), (\ref{init}). Then the variation constants formula
gives the following representation
for the solution of the problem (\ref{7.1}), (\ref{init}):
$$
 \left(\ba{c}q(0,t)\\ \dot q(0,t)\ea\right)= S(t)
 \left(\ba{c}q_0\\ p_0\ea\right)+\int_0^tS(\tau)
\left(\ba{c}0\\ z(1,t-\tau)\ea\right)\,d\tau,\quad t>0.
$$
Evidently, $S(0)=I$.
Moreover, the matrix $S(t)$ has a form
$\left(\ba{cc}\dot N(t)+\gamma N(t)& N(t)\\
\ddot N(t)+\gamma \dot N(t)&\dot N(t)\ea\right)$.
By Theorem~\ref{l3.1},
$|S(t)|\le C(1+t)^{-3/2}$, and
 the solutions of  (\ref{7.1}) satisfy the following bound:
 \begin{equation}\label{est-cs}
 |q(0,t)|+|\dot q(0,t)|\le C_1(1+t)^{-3/2}(|q_0|+|p_0|)
   +C_2\int_0^t(1+s)^{-3/2}|z(1,t-s)|\,ds,\quad t\ge0.
 \end{equation}
\end{cor}

%%%%%%%%%%%%%%%%%%%%%%%  Section 4   %%%%%%%%%%%%%%%%%%%%
\setcounter{equation}{0}
\section{Asymptotic behavior of $Y(t)$ as $t\to\infty$ in mean}  \label{sec5}
%%%%%%%%%%%%%%%%%%%%%%%%%%%%%%%%%%%%%
Set
$q^{(0)}(x,t)=q(x,t)$, $q^{(1)}(x,t)=\dot q(x,t)$, $x\in\Z_+$.
%%%%%%%%%%%%%%%%%%%%%%%%%%%%%% 4.1   %%%%%%%%%%%%%%%%%%%%
\subsection{The estimates of $q(0,t)$ in mean}
%%-----------------------------------------------

To derive the asymptotic behavior for $q^{(j)}(0,t)$ in mean,
introduce the following notations.
Write (see (\ref{3.2}))
\beqn\label{2.26}
{\bf G}^i_{z}(y,t):=\Big(
{\cal G}^{i0}_{t,+}(z,y),{\cal G}^{i1}_{t,+}(z,y)\Big)=
\Big({\cal G}_t^{i0}(z-y)-{\cal G}_t^{i0}(z+y),
{\cal G}_t^{i1}(z-y)-{\cal G}_t^{i1}(z+y)\Big),
\eeqn
$y,z\in\Z$, $i=0,1$, $t\in\R$.
%%%----------------------
Let ${\bf G}^j(y)$ denote the vector valued functions
 defined as
\beqn\label{Phi-j}
{\bf G}^j(y)=\int_0^{+\infty} N(s) {\bf G}_1^j(y,-s)\,ds=
\int_0^{+\infty} N^{(j)}(s) {\bf G}_1^0(y,-s)\,ds,\quad
y\in\Z, \quad j=0,1,
\eeqn
where the function $ N(s)$ is introduced in (\ref{N}),
 ${\bf G}_1^j(y,s)$ is defined in (\ref{2.26}).
%%------------------------------------------------------------------------------
\begin{pro} \label{pro-q}
Let conditions {\bf S1}--{\bf S4} and {\bf C} hold,
 $Y_0\in{\cal H}_{\alpha,+}$, and
 $q(0,t)$ be a solution of the problem (\ref{7.1}), (\ref{init}).
 Then
\be\label{5.4}
q^{(j)}(0,t)= \langle U_0(t) Y_0,{\bf G}^j\rangle_+ +r_j(t),\quad t>0,\quad j=0,1,
\quad \mbox{where }\, \E|r_j(t)|^2\le C(1+t)^{-1},
\ee
the vector valued functions ${\bf G}^j$ are defined in (\ref{Phi-j}).
\end{pro}
%%%%%%%%%%%%%%%%%%%%%%%%%%%%%%%%%%%%%%%%%%%%%%%%%%%%%%%
{\bf Proof}\, First, Corollary \ref{cor3} and the bound (\ref{NN}) imply that
$$
\E\left|\left(\ba{c}q(0,t)\\ \dot q(0,t)\ea\right)
-\int_0^t S(\tau)\left(\ba{c}0\\z(1,t-\tau)\ea
\right)d\tau \right|^2\le C(1+t)^{-3},\quad t>0.
$$
Second,
$S(\tau)\left(\ba{c}0\\ z(1,t-\tau)\ea\right)=
\left(\ba{c} N(\tau)\\\dot N(\tau)\ea\right) z(1,t-\tau)$.
Moreover, for $j=0,1$,
$$
\E\left|\int\limits_t^{+\infty}
 N^{(j)}(\tau) z(1,t-\tau)d\tau \right|^2=
\int\limits_t^{+\infty} N^{(j)}(\tau_1)d\tau_1
\int\limits_t^{+\infty} N^{(j)}(\tau_2)\,
\E\Big(z(1,t-\tau_1)z(1,t-\tau_2)\Big)d\tau_2.
$$
  Further, the bound (\ref{ubound}) gives
$$
\left|\E\left(z(1,t-\tau_1)z(1,t-\tau_2)\right)\right|\le C
\sup_{s\in\R}\E|z(1,s)|^2\le C_1<\infty.
$$
Hence, the bound (\ref{NN}) yields
$$
\E\left| \int_t^{+\infty}
S(\tau)\left(\ba{c}0\\ z(1,t-\tau)\ea \right)d\tau
\right|^2\le
C\sup_{s\in\R}\E|z(1,s)|^2\left(\int_t^{+\infty}(1+\tau)^{-3/2}d\tau\right)^2
\le C_1(1+t)^{-1}.
$$
This implies the representation (\ref{5.4}), since by (\ref{2.26}) and (\ref{sol}),
$$
 z(1,t-\tau)=\langle U_0(t)Y_0(\cdot), {\bf G}_1^0(\cdot,-\tau)\rangle_+.
 \bo
$$
%%%%------------------------------------------
\begin{remark}\label{rem2.10}
{\rm (i) By (\ref{3.2}) and (\ref{hatcalG}),
${\bf G}^i_{z}(y,t)$ is odd w.r.t. $y\in\Z$.
Also, the Parseval identity  gives
\beqn\label{2.27}
\Vert{\bf G}^i_{z}(\cdot,t)\Vert^2_{0}= \frac1{2\pi}
\int_{-\pi}^{\pi} |\hat{\bf G}^i_{z}(\theta,t)|^2\,d\theta
= \frac2\pi\int_{-\pi}^\pi
\Big( |\hat{\cal G}^{i0}_t(\theta)|^2
+|\hat{\cal G}^{i1}_t(\theta)|^2\Big)\sin^2(z\theta)
\,d\theta,\,\,\,\,  z\in\Z.
\eeqn
Hence, if $m\not=0$, then $\Vert{\bf G}^i_{z}(\cdot,t)\Vert^2_{0}\le C<\infty$
uniformly in $z\in\Z$ and $t\in\R$.
If $m=0$, then $\Vert{\bf G}^1_{z}(\cdot,t)\Vert^2_{0}\le C<\infty$ uniformly on $z$ and $t$,
and
$$
\Vert{\bf G}^0_{z}(\cdot,t)\Vert^2_{0}
\le C_1+C_2\int_\pi^\pi\frac{\sin^2(z\theta)}{\sin^2(\theta/2)}\,d\theta\le C_1+C_2|z|,
\quad z\in\mathbb{Z}.
$$
In particular,
$\hat {\bf G}_1^0(\theta,t)
=2i\sin\theta\Big(\cos(\phi(\theta)t), \sin(\phi(\theta)t)/\phi(\theta)\Big)$,
$\theta\in\T$.
Hence,
\be\label{2.28}
\Vert{\bf G}^0_{1}(\cdot,t)\Vert^2_{0}=C
\int_{-\pi}^\pi
\Big( \cos^2(\phi(\theta) t)
+\frac{\sin^2(\phi(\theta)t)}{\phi^2(\theta)}\Big)\sin^2(\theta)\,d\theta\le C<\infty.
\ee
%%%------------------------------------------------------
(ii) ${\bf G}^j(\cdot)$ is odd. Moreover, by the bounds (\ref{NN}) and (\ref{2.28}),
 ${\bf G}^j\in{\cal H}_0$, since
\be\label{6.5'}
 \Vert{\bf G}^j(\cdot)\Vert_{0}\le
 \int_0^{+\infty}|N^{(j)}(s)|\Vert{\bf G}_1^0(\cdot,-s)\Vert_{0}\,ds
 \le C\int_0^{+\infty}|N^{(j)}(s)|\,ds<\infty.
 \ee
In Fourier transform,
\be\label{4.7}
\hat {\bf G}^j(\theta)
=2i\sin\theta\int_0^{+\infty}
N^{(j)}(s)\Big(\cos(\phi(\theta)s), -\sin(\phi(\theta)s)/\phi(\theta)\Big)ds,
\quad \theta\in\T.
\ee
Therefore,
${\bf G}^j\in{\cal H}_{1/\phi}:=\ell^2_{1/\phi}\times\ell^2$, where
$\ell^2_{1/\phi}=\{\psi\in\ell^2: \hat{\psi}/\phi\in L^2(\T)\}$.
Note that $\ell^2_{1/\phi}\equiv\ell^2$ if $m\not=0$.
}\end{remark}
%%-------------------------

Denote by $U'_0(t)$ the operator adjoint to $U_0(t)$:
\be\label{defU'}
\langle Y,U'_0(t)\Psi\rangle_+=\langle U_0(t)Y,\Psi\rangle_+,\quad
Y\in{\cal H}_{\alpha,+},\quad \Psi\in{\cal S},\quad t\in\R.
\ee
In other words,
$$
(U'_0(t)\Psi)^j(y)=\sum\limits_{i=0,1}\sum\limits_{x\in\Z_+}
{\cal G}^{ij}_{t,+}(x,y)\Psi^i(x)
\quad \mbox{for }\,\,\Psi=(\Psi^0,\Psi^1)\in{\cal S},\,\,\, t\in\R,\,\,\, y\in\Z_+,\,\,\, j=0,1.
$$
In particular,
${\bf G}^0_{1}(y,t)=(U'_0(t) \Psi)(y)$ with $\Psi(x)=(\delta_{1x},0)$ (see (\ref{2.26})),
where $\delta_{1x}$ denotes the Kronekker symbol.
%%%%%%%%%%%%%%%%%%%%%%%%%%%%%%%%%%%%%%%%%
\begin{cor}
  (i) For ${\bf G}^j$ defined in (\ref{Phi-j}), we have
  \be\label{6.6}
  \sup\limits_{t\in\R}\E|\langle U_0(t)Y_0,{\bf G}^j\rangle_+|^2=
  \sup\limits_{t\in\R}\E|\langle Y_0,U'_0(t){\bf G}^j\rangle_+|^2\le C<\infty.
  \ee
  Indeed, since $U'_0(t){\bf G}_1^0(y,\tau)={\bf G}_1^0(y,\tau+t)$,  we have
\be\label{6.66}
\sup_{t\in\R}\Vert U'_0(t){\bf G}^j\Vert_0\le C<\infty,
\ee
by (\ref{Phi-j}), (\ref{2.28}) and (\ref{NN}).
Hence, (\ref{6.6}) follows from the bounds (\ref{2.42}) and (\ref{6.66}).
\smallskip\\
 (ii) The representation (\ref{5.4}), Lemma \ref{l1} and the bound (\ref{6.6}) give
 the following convergence,
\be\label{6.22}
\E\Big(q^{(i)}(0,t)q^{(j)}(0,t)\Big)\to {\cal Q}^{\nu}_\infty({\bf G}^i,{\bf G}^j)
\quad \mbox{as }\, t\to\infty,
\ee
where the quadratic form  $ {\cal Q}^{\nu}_\infty$ is defined by (\ref{qpp}).
The r.h.s. of (\ref{6.22}) is defined by Remarks~\ref{rem2.8} and \ref{rem2.10} (ii).
%%%%%%%%%%%%%%%%%%%%%%%%%%%%%%%% add
Furthermore, using formulas  (\ref{2.24}), (\ref{4.7}), and (\ref{q-infty}),
we obtain ${\cal Q}^{\nu}_\infty({\bf G}^i,{\bf G}^j)=0$ for $i\not=j$, $i,j=0,1$, and
\beqn\label{4.10'}
\lim_{t\to\infty}\E\left( q^{(i)}(0,t)\right)^2={\cal Q}^{\nu}_\infty({\bf G}^i,{\bf G}^i)=
\frac{2}{\pi}\int_{\T}\sin^2(\theta)\,\hat q^{ii}_\infty(\theta)
\Big|\int_0^{+\infty}N(s)e^{i\phi(\theta)s}\,ds\Big|^2 d\theta.
\eeqn
\end{cor}

%%%%%%%%%%%%%%%%%%%%%%%%%% 4.2  %%%%%%%%%%%%%%%%%%%%%%%%%%%%%%%%%
\subsection{The large time behavior of $q(x,t)$, $x\in\N$, in mean}
%%%%--------------------------------------------------

Let ${\bf K}^{j}(x,y)$  $j=0,1$, $x\in\N$, $y\in\Z$, stand for  vector-valued functions of a form
\beqn\label{Pi-0}
{\bf K}^j(x,y)=
\int\limits_0^{+\infty}K(x,s)\Big(U'_0(-s){\bf G}^j\Big)(y)\,ds
=\int\limits_0^{+\infty}\int\limits_0^{+\infty}
K(x,s) N^{(j)}(\tau){\bf G}^0_1(y,-s-\tau)\,ds\,d\tau,
\eeqn
where  $K(x,s)$ is defined in (\ref{K(x,t)}),
${\bf G}^j$ is introduced in (\ref{Phi-j}).
Note that ${\bf K}^j(x,y)$ is odd w.r.t. $y\in\Z$,
${\bf K}^j(x,\cdot)\in{\cal H}_{0}$ for any $x\in\N$
and $\Vert{\bf K}^j(x ,\cdot)\Vert_0\in{\cal H}_{\alpha,+}$ with $\alpha<-3/2$
by (\ref{boundK}) and (\ref{6.66}).

%%-------------------------------------
\begin{lemma} \label{pro-qx}
Assume that $\alpha<-3/2$. Then the following assertions are true.\\
(i) The solution $q(x,t)$, $x\in\N$, of the problem (\ref{b.1})--(\ref{b.4})
admits the following representation
\be\label{qj}
q^{(j)}(x,t)
= \langle U_0(t) Y_0,{\bf K}^j(x,\cdot)\rangle_+ +r_j(x,t),
\quad j=0,1,\quad t>0,
\ee
where $\E(\Vert r_j(\cdot,t)\Vert'_{\alpha,+})^2\le C\langle t\rangle^{-1}$,  %%%${\bf K}^j$ is introduced in (\ref{Pi-0}).
and the notation $\Vert \cdot\Vert'_{\alpha,+}$ is introduced in (\ref{boundK}).
\smallskip\\
(ii)
The correlation functions  converge to a limit,
$$
\lim_{t\to\infty}
\E\Big(q^{(i)}(x,t)q^{(j)}(x',t)\Big)
={\cal Q}^\nu_\infty({\bf K}^i(x,\cdot),{\bf K}^j(x',\cdot))
\quad \mbox{for any }\,\,x,x'\in\Z_+,\quad i,j=0,1.
$$
\end{lemma}
%%%%---------------------------
{\bf Proof}\,
At first, by (\ref{qxt}) and (\ref{5.4}),
\be\label{6.1}
q^{(j)}(x,t)=\int_{0}^t K(x,t-s)q^{(j)}(0,s)\,ds=
\int_{0}^t K(x,t-s)\langle U_0(s)Y_0,{\bf G}^j(\cdot)\rangle_+\,ds
+r'_j(x,t),
\ee
where $x\in\N$, $\E\Vert r'_j(\cdot,t)\Vert^2_{\alpha,+}\le C\langle t\rangle^{-1}$.
Indeed, by (\ref{5.4}) and (\ref{boundK}),
\beqn
\E\left(\Vert r'_j(\cdot,t)\Vert'_{\alpha,+}\right)^2\!\!\!\!&=&\!\!
\E\Big(\Big\Vert \int_{0}^t K(\cdot,t-s) r_j(s)\,ds\Big\Vert'_{\alpha,+}\Big)^2\le
\Big(\int_0^t \Vert K(\cdot,t-s)\Vert'_{\alpha,+}\sqrt{\E|r_j(s)|^2}\,ds\Big)^2
\nonumber\\
\!\!\!\!&\le&\!\! C\Big(\int_0^t (1+t-s)^{-3/2}(1+s)^{-1/2}\,ds\Big)^2\le C_1\langle t\rangle^{-1}.
\nonumber
\eeqn
Second, the first term in the r.h.s. of (\ref{6.1}) has a form (see (\ref{Pi-0}))
\be\label{4.14}
\int_{0}^t K(x,s)\langle U_0(t-s)Y_0,{\bf G}^j(\cdot)\rangle_+\,ds=
\langle U_0(t)Y_0,{\bf K}^j(x,\cdot)\rangle_++r''_j(x,t),
\ee
where, by definition,
$$
 r''_j(x,t)=\displaystyle\int_t^{+\infty}K(x,s)
\langle U_0(t-s)Y_0,{\bf G}^j\rangle_+\,ds.
$$
The bounds (\ref{boundK}) and (\ref{6.6}) yield
\beqn\label{6.5}
\E\left(\Vert r''_j(\cdot,t)\Vert'_{\alpha,+}\right)^2\le\Big(\int_t^{+\infty}
\Vert K(\cdot,s)\Vert'_{\alpha,+}
\Big(\E_0\left|\langle U_0(t-s)Y_0,{\bf G}^j\rangle_+\right|^2\Big)^{1/2}\,ds\Big)^2
\le C\langle t\rangle^{-1}.
\eeqn
Hence, the bounds (\ref{6.1})--(\ref{6.5}) imply (\ref{qj})
with $r_j(x,t)=r'_j(x,t)+r''_j(x,t)$.
Finally, the assertion~(ii) of the lemma follows from the representation
(\ref{qj}) and Lemma \ref{l1}.\bo
\medskip\\
%%%%%%%%%%%%%%%%%%%%%%%%%%%%%%%
{\bf Remark}\, {\em Set
$\tilde K(0,\omega):=e^{i\theta(\omega)x}|_{x=0}=1$. Then, formally, $K(0,t)=\delta_{0t}$.
Hence, we can put   ${\bf K}^j(0,y)={\bf G}^j(y)$, $y\in\Z$.
Then the representation (\ref{5.4}) follows from (\ref{qj}).}
\medskip
%%%%%%%%%%%%%%%%%%%%%%%%%%%%%%%%

For any $\psi\in S$ and $j=0,1$, denote
\be\label{Kpsi}
{\bf K}^j_\psi(y):=
\langle {\bf K}^j(\cdot,y),\psi(\cdot)\rangle_+=
\int_0^{+\infty}
\langle K(\cdot,s),\psi\rangle_+ (U'_0(-s){\bf G}^j)(y)\,ds,
\quad y\in\Z.
\ee
%%-----------------------------------
\begin{remark}\label{rem6.6}
By Remark \ref{rem2.10} and (\ref{Pi-0}),
  ${\bf K}^j(x,y)$ is odd w.r.t. $y\in\Z$.
Then,  ${\bf K}^j_\psi$ is odd.
Moreover, ${\bf K}^j(x,\cdot)\in{\cal H}_{1/\phi}\equiv\ell^2_{1/\phi}\times\ell^2$
for any $x\in\Z_+$.
Therefore, ${\bf K}^j_\psi\in{\cal H}_{1/\phi}$.
\end{remark}
%%%%%%%%%%%%%%%%%%%%%%%%%%%%%%5
\begin{cor}\label{cor6}
(i) For any $\psi\in S$ and $j=0,1$,
\be\label{6.15}
\langle q^{(j)}(\cdot,t),\psi\rangle_+
=\langle U_0(t) Y_0,{\bf K}^j_\psi\rangle_+ +r_j(t),\quad t>0,
\quad \mbox{where }\,\,\E|r_j(t)|^2\le C\langle t\rangle^{-1}.
\ee
(ii) For any $\psi\in S$,
\be\label{6.14}
\sup_{t\in\R}\E|\langle U_0(t)Y_0,{\bf K}^j_\psi\rangle_+|^2=
\sup_{t\in\R}\E|\langle Y_0,U'_0(t){\bf K}^j_\psi\rangle_+|^2\le C<\infty.
\ee
(iii) For any $\psi,\chi\in S$,
$
\E\Big(\langle q^{(i)}(\cdot,t),\psi\rangle_+\langle q^{(j)}(\cdot,t),\chi\rangle_+\Big)
\to {\cal Q}_\infty^{\nu}({\bf K}^i_{\psi},{\bf K}^j_{\chi})$
as $t\to\infty$.
\end{cor}
%%%%%%%%%%%%%%%%%%%%%%%%%%%%%%5
{\bf Proof}\,
The representation (\ref{6.15}) follows from (\ref{qj}).
Now we verify  (\ref{6.14}). By  (\ref{Kpsi}),
$$
U'_0(t){\bf K}^j_\psi=\int_0^{+\infty}
\langle K(\cdot,s),\psi\rangle_+ U'_0(t-s) {\bf G}^j\,ds.
$$
Hence, (\ref{6.66}) and (\ref{boundK}) imply the following uniform bound,
\be\label{6.17}
\sup_{t\in\R}
\Vert U'_0(t){\bf K}^j_\psi\Vert_0
\le C\int_0^{+\infty}\Vert K(\cdot,s)\Vert_{\alpha,+}\,ds<\infty.
\ee
Therefore,
(\ref{6.14}) follows from the bounds (\ref{2.42}) and (\ref{6.17}).
Finally, (\ref{6.15}) and Lemma \ref{l1} imply the part (iii) of Corollary \ref{cor6}.
%Note that all assertions of Corollary \ref{cor6} remain true if $\psi\in\ell^2_{-\alpha,+}$
%with $\alpha<-3/2$.
\bo\\
\smallskip

%%-----------------------------------------------------------------
Write $\Pi^j(x,y)=e_{j}\delta_{xy}+{\bf K}^{j}(x,y)$ with
$e_0=(1,0)$, $e_1=(0,1)$, $j=0,1$, $x,y\in\Z_+$.
Given $\Psi=(\Psi^0,\Psi^1)\in{\cal S}$, define the vector valued functions
$\Pi_\Psi(y)$, $y\in\Z_+$,
\be\label{hn}
\Pi_\Psi(y):=\sum\limits_{j=0}^1\langle \Pi^j(\cdot,y),\Psi^j\rangle_+=
\Psi(y)+\sum\limits_{j=0}^1
\langle {\bf K}^{j}(\cdot,y),\Psi^j(\cdot)\rangle_+
=\Psi(y)+\sum\limits_{j=0}^1{\bf K}^j_{\Psi^j}(y)
\ee
with ${\bf K}^j_{\psi}(y)$ from (\ref{Kpsi}).
The formula (\ref{2.1}) and Lemma \ref{pro-qx}
imply the following lemma.
 %%---------------------------------------
\begin{lemma}\label{l6.7}
For the solution $Y(t)=(u(x,t),\dot u(x,t))\equiv (Y^0(x,t), Y^1(x,t))$
of the problem (\ref{1.1})--(\ref{1.3}),
 the following asymptotics holds,
$$
Y^j(x,t)= \langle U_0(t) Y_0,\Pi^j(x,\cdot)\rangle_+ +r_j(x,t),\quad j=0,1,
$$
where $r_j(x,t)$ is introduced in (\ref{qj}) and
$\E\Vert r_j(\cdot,t)\Vert_{\alpha,+}^2\le C\langle t\rangle^{-1}$ ($\alpha<-3/2$). Hence,
for any $\Psi=(\Psi^0,\Psi^1)\in{\cal S}$,
\be\label{6.21}
\langle Y(t),\Psi\rangle_+=\langle U_0(t)Y_0,\Pi_\Psi\rangle_++r(t),
\ee
where $r(t)=\sum_{j=0}^1\langle r_j(\cdot,t),\Psi^j\rangle_+$ and
$\E|r(t)|^2\le C\langle t\rangle^{-1}$.
\end{lemma}

%%%%%%%%%%%%%%%%%%%%%%%%%%%%% Section 5   %%%%%%%%%%%%%%%%%%%%%%%%%%%
\setcounter{equation}{0}
\section{Proof of Theorem \ref{t0}}   \label{s5}
%%%%%%%%%%%%%%%%%%%%%%%%%%%%%%%%%%%

The compactness of the measures family $\{\mu_t,\,t>0\}$
follows from the Prokhorov compactness theorem and the bound (\ref{c.1}) below,
since the embedding ${\cal H}_{\alpha,+}\subset{\cal H}_{\beta,+}$
is compact if $\alpha>\beta$.
%%%%%%%%%%%%--------------------------------------------------------------------------
\begin{lemma} Let $\alpha<-3/2$ and conditions {\bf S1}--{\bf S4} and {\bf C} hold.
 Then
\be\label{c.1}
\sup_{t\ge0}\E\Vert U(t)Y_0\Vert^2_{\alpha,+}\le C<\infty.
\ee
\end{lemma}
{\bf Proof}\,
By  (\ref{2.1}), the solution $Y(\cdot,t)=U(t)Y_0$
 has a form $Y(x,t)=Z(x,t)+X(x,t)$,
 where $Z(x,t)=U_0(t)Y_0$ and $X(x,t)=(q(x,t),\dot q(x,t))$.
 Hence, by the bound (\ref{ubound}),
 to prove the bound (\ref{c.1}) it suffices to verify that
 \be\label{6.2}
 \sup_{t\ge0}\E\Vert X(\cdot,t)\Vert^2_{\alpha,+}\le C<\infty.
\ee
Applying  (\ref{qxt}) and  (\ref{boundK}), we have
\beqn\label{6.3}
\E\Vert X(\cdot,t)\Vert^2_{\alpha,+}&=&\E|X(0,t)|^2+
\E\sum\limits_{x\in\N}\langle x\rangle^{2\alpha}
\left|\int_0^tK(x,t-s)X(0,s)\,ds\right|^2
\nonumber\\
&\le& C\sup\limits_{\tau\in[0,t]}\E|X(0,\tau)|^2.
\eeqn
Using the estimate (\ref{est-cs}), we obtain
$$
\E|X(0,t)|^2\le \E(|q_0|^2+|p_0|^2)+C\sup_{\tau\ge0}\E|z(1,\tau)|^2,
\quad t\ge0.
$$
On the other hand,
$\sup\limits_{\tau\in\R}\E|z(1,\tau)|^2=\sup\limits_{\tau\in\R} Q^{00}_\tau(1,1)\le C<\infty$
by  (\ref{ubound}). Hence,
\be\label{6.4}
\sup_{\tau\ge0}\E|X(0,t)|^2\le C<\infty.
\ee
Estimates (\ref{6.3}) and (\ref{6.4}) imply the bound (\ref{6.2}).
\bo\\
\smallskip

 To end the proof of assertion~(i)
of Theorem \ref{t0}, it suffices to check the convergence (\ref{2.6i})
of characteristic functionals for $\mu_t$. By the triangle inequality
and the equality (\ref{Qmu}),
  \beqn\label{8.16}
   \left|\E e^{i\langle Y(t),\Psi\rangle_+}-
e^{-\frac{1}{2} {\cal Q}_\infty(\Psi,\Psi)}\right|
&\le&\Big|\E \Big(e^{i\langle Y(t),\Psi\rangle_+}-
 e^{i\langle U_0(t)Y_0,\Pi_\Psi\rangle_+}\Big)\Big|
\nonumber\\
&&+ \Big|\E_0\left( e^{i\langle U_0(t)Y_0,\Pi_\Psi\rangle_+}\right)
-e^{-\frac{1}{2}{\cal Q}^\nu_\infty (\Pi_\Psi,\Pi_\Psi)}\Big|.
 \eeqn
 Applying (\ref{6.21}), we estimate
 the first term in the r.h.s. of (\ref{8.16}) by
$$
   \E\Big|\langle Y(t),\Psi\rangle_+
-\langle U_0(t)Y_0,\Pi_\Psi\rangle_+\Big|
 \le \E|r(t)| \le
\Big(\E|r(t)|^2\Big)^{1/2} \le C \langle t\rangle^{-1/2}\to0\quad
\mbox{as }\,t\to\infty.
  $$
 The convergence of the characteristic functionals
$\hat\nu_t(\Pi_\Psi)=
\E_0\left(\exp\{i\langle U_0(t)Y_0,\Pi_\Psi\rangle_+\}\right)$ to a limit
as $t\to\infty$ follows from Theorem \ref{t1} and Remark \ref{rem2.11}.
This proves the convergence (\ref{2.6i}).
Corollary \ref{cor6} and Lemma \ref{l1} imply assertion~(ii) of Theorem \ref{t0}.

%%%%%%%%%%%%%%%%%%%%%%%%%%%%%%%%%%%%%%%%%%%%%%%%
\medskip\medskip

{\bf ACKNOWLEDGMENTS}
\medskip

This work was supported partly by the research grant of RFBR
(Grant No. 15-01-03587).
The author is grateful to
Herbert Spohn for the statement of the problem and to
Alexander Komech for useful
discussions concerning several aspects of this paper.

%%%%%%%%%%%%%%%%%%%%%%%%%%%%%%%%%%%   Appendix A  %%%%%%%%%%%%%%%%
\appendix
\setcounter{section}{1}
\setcounter{equation}{0}
\setcounter{theorem}{0}
\section*{\large\bf Appendix A: The behavior of $K(x,t)$ as $t\to\infty$}
%%%%%%%%%%%%%%%%%%%%%%%%%%%%%%%%%%%%%%%%%%%%%%%%%%%%%%%%

In this section, we first study the  properties of $\tilde K(x,\omega):=e^{i\theta(\omega)x}$
(the function $\theta(\omega)$ is introduced in Lemma~\ref{theta})
applying the technique of  \cite{KKK, D16}.
Next, using these properties, we obtain the bound (\ref{boundK}) for $K(x,t)$.
Finally, by the bound (\ref{boundK}), we prove the existence of solutions of the problem (\ref{1.1})--(\ref{1.3}).

%%%%%%%%%%%%%%%%%%%%%%%%%%  A.1    %%%%%%%%%%%%%%%%%5
\subsection{Properties of $e^{i\theta(\omega)x}$
for $\omega\in\mathbb{C}$ and  $x\in\N$ }
%%%---------------------------

We set $\Lambda:=[-\sqrt{4+m^2},-m]\cup[m,\sqrt{4+m^2}]$ and $\Lambda_0=\{\pm m, \pm \sqrt{4+m^2}\}$.
We indicate the properties {\bf(I)}--{\bf(III)} of the function $ \tilde K(x,\omega)$
for different values of $\omega$ such as $\omega\in \mathbb{C}\setminus \Lambda$,
$\omega\in \Lambda\setminus\Lambda_0$, and $\omega\in \Lambda_0$.
\smallskip\\
{\bf (I)}\, Let $\omega\in \mathbb{C}\setminus \Lambda$. Then $\Im\theta(\omega)>0$.
In this case,
$\tilde K(x,\omega)$ exponentially decays in $x$. Hence,  $\tilde K(x,\omega)$
is an analytic function in the complex  $\omega$-plane
with the values in the class $\ell^2_{\alpha,+}$.
Moreover, by (\ref{32}) and the condition $\Im\theta(\omega)>0$, we have
\be\label{7.0}
\left| e^{i\theta(\omega)}\right|\le C|\omega|^{-2}\quad \mbox{as }\,\,|\omega|\to\infty.
\ee
%%------------------------------------------------------------------------------------
%Further, we note that if $\Im\omega>0$, then
%$\Im\theta(\omega)\to+\infty$ and $\Re\theta(\omega)\to\pm\pi$ as $\Re\omega\to\pm\infty$.
%(If $\Im\omega<0$, then $\Im\theta(\omega)\to+\infty$ and $\Re\theta(\omega)\to\mp\pi$
%as $\Re\omega\to\pm\infty$.)
%%%%$\Im \theta(\omega)\sim \ln(|\omega|^2)$ as $|\omega|\to\infty$
%%Secondly, for any $K>0$ there exists a positive constant $C=C(K)<\infty$ such that
%%\be\label{7.0} \left|e^{i\theta(\omega)}\right|\le C|\Re\omega|^{-2}
%%\quad \mbox{as }\,\, \Re \omega\to\infty \ee for any $\Im\omega\in\R$: $|\Im\omega|\le K$.
%%Indeed, put $z=1-(\omega^2-m^2)/2$. Then, by (\ref{32}),
%%\be\label{7.2} e^{i\theta(\omega)}=\cos\theta+i\sin\theta=z+\sqrt{z^2-1},%%%=\frac1{z-\sqrt{z^2-1}}\ee
%%where $\sqrt{z^2-1}$ is a complex root and its branch is chosen by the condition $\Im\theta(\omega)>0$.
%In particular, this condition implies that ${\rm sgn}(\Im\sqrt{z^2-1})=-{\rm sgn} (\Im z)=
%{\rm sgn}(\Im\omega^2)$ for $\Im\omega^2\not=0$.
%Moreover,   $\sqrt{z^2-1}>0$ for $\omega\in(-\infty,-\sqrt{4+m^2})\cup(\sqrt{4+m^2},+\infty)$.
%Let us verify (\ref{7.0}) for $\omega=\nu\in\R$: $|\nu|>\sqrt{4+m^2}$.
%For such  $\omega$, $\Re(\theta(\nu))=\pm\pi$. Hence, by (\ref{32}),
%$$e^{i\theta(\nu)}=-e^{-\Im\theta(\nu)}=\frac2{2-(\nu^2-m^2)-\sqrt{(2-\nu^2+m^2)^2-4}}
%\sim -\frac1{\nu^2} \quad \mbox{as }\, |\nu|\to\infty.$$
%In the general case, the bound (\ref{7.0}) can be proved similarly.
%%%-------------------------------------------------
\begin{remark}
It follows from (\ref{K(x,t)}) and (\ref{7.0})  that there exist constants $C,B<\infty$ such that
\be\label{7.4}
\sum\limits_{x\in\N} \langle x\rangle^{2\alpha}|K(x,t)|^2\le Ce^{Bt}
\quad \mbox{for any }\,\alpha\in\R,\quad t>0.
\ee
\end{remark}
%%-------------------------------------------------------------------------------------------
%%{\bf Proof}\, Let us apply the inverse Fourier--Laplace transform to $\tilde K(x,\omega)$.
%%Then for some $\mu>0$, any $x\in\N$, and $t>0$, we have
%%$$ K(x,t)=\frac1{2\pi}\int\limits_{-\infty+i\mu}^{+\infty+i\mu}
%%e^{-i\omega t}e^{i\theta(\omega)x}\,d\omega
%%=\frac{e^{\mu t}}{2\pi}\int\limits_{-\infty}^{+\infty}e^{-i\nu t}
%%e^{i\theta(\nu+i\mu)x}\,d\nu,$$
%%where $\omega=\nu+i\mu$ ($\mu,\nu\in\R$). Note that
%%$\left| e^{i\theta(\nu+i\mu)}\right|=e^{-\Im \theta(\nu+i\mu)}<1$
%%for any $\nu\in\R$ and $\mu>0$.
%Hence, $$ \sum\limits_{x\in\N} \langle x\rangle^{2\alpha}|K(x,t)|^2\le
%\frac{e^{2\mu t}}{4\pi^2}\int\limits_{-\infty}^{+\infty}
%\sum\limits_{x\in\N} \langle x\rangle^{2\alpha}\left(\delta^2_{\nu,\mu}\right)^x\,d\nu.$$
%%%%-------------------------------------------------------------
{\bf (II)}\, Let $\omega\in\Lambda\setminus \Lambda_0$.
Then we have the pointwise convergence
 $ \tilde K(x,\omega\pm i\ve)\to \tilde K(x,\omega\pm i0)$ as $\ve\to+0$
 for any $x\in\N$.
Moreover, $|\tilde K(x,\omega\pm i\ve)|\le C$ uniformly in $\varepsilon$.
Hence, for any $\alpha<-1/2$ and $\omega\not\in\Lambda_0$,
$$
\sum\limits_{x\in\N}\langle x\rangle^{2\alpha}
 |\tilde K(x,\omega\pm i0)-\tilde K(x,\omega\pm i\ve)|^2\to0\quad \mbox{as }\,\,\ve\to+0,
$$
 by the Lebesgue dominated convergence theorem.
Furthermore, since $\overline{\theta(\omega)}=-\theta(\bar \omega)$
for $\omega\in \mathbb{C}\setminus \Lambda$,
$\tilde K(x,\omega-i0)=\overline{\tilde K(x,\omega+i0)}$
for $\omega\in\Lambda\setminus \Lambda_0$ and $x\in\N$.
\smallskip\\
%%%-----------------------------------------------------------------------------------
{\bf (III)}\, Now we study the behavior of $\tilde K(x,\omega)$ near
the  points $\omega\in\Lambda_0=\{\pm m,\pm\sqrt{4+m^2}\}$.
Eqn~(\ref{32})  implies
\be\label{a3}
e^{i\theta(\omega)}=\cos\theta(\omega)+i\sin\theta(\omega)
=1-\frac12(\omega^2-m^2)+\frac{i}2
\sqrt{(\omega^2-m^2)(4+m^2-\omega^2)},
\quad \omega\in \mathbb{C}\setminus\Lambda.
\ee
The Taylor expansion implies
\be\label{a8}
e^{i\theta(\omega)}= 1+i\sqrt{\omega^2-m^2}-
\frac12(\omega^2-m^2)-\frac{i}8(\omega^2-m^2)^{3/2}+\dots,
\quad \omega\to \pm m,
\ee
where $\omega\in\mathbb{C}_+:=\{\omega\in \mathbb{C}:\Im\omega>0\}$,
$\Im\sqrt{\omega^2-m^2}>0$.
Here ${\rm sgn}(\Re\sqrt{\omega^2-m^2})={\rm sgn}(\Re\omega)$
for $\omega\in \mathbb{C}_+$.
 This choice of the branch of the complex root  $\sqrt{\omega^2-m^2}$ follows
 from the condition $\Im\theta(\omega)>0$.
Hence, for $x\in\N$,
\be\label{a5}
e^{i\theta(\omega)x}=1+\sum\limits_{j=1}^{+\infty}
(\omega^2-m^2)^{j/2} R^j(x),\quad \omega\to \pm m+i0.
\ee
Here polynomials $R^j(x)$
are of a form  $R^j(x)=\sum\limits_{k=1}^j c^j_k x^k$
with  coefficients $c_k^j\in\mathbb{C}$, $j\in\N$.
For example, $R^1(x)=ix$, $R^2(x)=-x^2/2$, $R^3(x)=-i(4x^3-x)/24$.

Similarly,
\be\label{a9}
e^{i\theta(\omega)}=-1+i\sqrt{m^2+4-\omega^2}+\frac12(m^2+4-\omega^2)
-\frac{i}8(m^2+4-\omega^2)^{3/2}+\dots
\ee
as $\omega\to\pm\sqrt{m^2+4}$, $\omega\in \mathbb{C}_+$.
Here $\sqrt{m^2+4-\omega^2}$ is the complex root and we choose the branch of this root
so that ${\rm sgn}(\Re\sqrt{m^2+4-\omega^2})={\rm sgn}(\Re\omega)$,
which follows from the condition $\Im\theta(\omega)>0$.
 Hence, for $x\in\N$,
 \beqn\label{a6}
 e^{i\theta(\omega)x}=
 (-1)^x\left(1-ix\sqrt{m^2+4-\omega^2}-(m^2+4-\omega^2)x^2/2+\dots\right)
\eeqn
as $\omega\to \pm\sqrt{m^2+4}+i0$.
If $m=0$, then (\ref{a3}) and the Taylor expansion give
\be\label{a10}
e^{i\theta(\omega)}=
%1-\frac{\omega^2}{2}+i\sqrt{\omega^2-\frac{\omega^4}{4}}=
1-\frac{\omega^2}{2}+i\omega(1-\frac{\omega^2}8-\frac{\omega^4}{128}+\dots)
\quad \mbox{as }\,\,\omega\to0,
\ee
and $e^{i\theta(\omega)}= -1+i\sqrt{4-\omega^2}+\dots$ as $\omega\to \pm 2$, $\omega\in \mathbb{C}_+$.
Therefore, in the case $m=0$,
\be\label{a7}
e^{i\theta(\omega)x}=\left\{\ba{ll}
1+i\omega x-\omega^2x^2/2-i\omega^3(4x^3-x)/24 +\dots& \mbox{as  }\,\omega\to0\\
(-1)^x(1-ix\sqrt{4-\omega^2}-(4-\omega^2)x^2/2+\dots)
& \mbox{as  }\, \omega\to\pm2+i0
\ea\right| x\in\N.
\ee
The representations (\ref{a5}), (\ref{a6}) and (\ref{a7}) lead to the following result.
%%%%%%%%%%%%%%%%%%%%%%%%%%%%%%%%%%%%%%%%%%%%%%%%%%%%%%%%%
\begin{lemma} (cf \cite[Lemma 3.2]{KKK})
For every $N\in\N$,
\be\label{3.8}
e^{i\theta(\omega)x}=1+\sum\limits_{j=1}^N
(\omega^2-m^2)^{j/2} R^j(x)+r_N(\omega,x),\quad \omega\to\pm m+i0,
\ee
where
$\Vert r_N(\omega,\cdot)\Vert_{\alpha,+}
= O(|\omega^2-m^2|^{(N+1)/2})$ for $\alpha<-3/2-N$.
Moreover,
$$
D^r_\omega\left(e^{i\theta(\omega)x}\right)
=\sum\limits_{j=1}^N
\frac{d^r}{d\omega^r}(\omega^2-m^2)^{j/2} R^j(x)+\tilde r_N(\omega,x),
\quad \omega\to\pm m+i0,
$$
where
$\Vert \tilde r_N(\omega,\cdot)\Vert_{\alpha,+}
= O(|\omega^2-m^2|^{(N+1)/2-r})$ for $\alpha<-3/2-N$.
The similar representation holds for $\omega\to\pm\sqrt{m^2+4}+i0$.
\end{lemma}
%%------------------

Indeed, the bound (\ref{3.8}) follows from the following representation for
remainder (see formula (\ref{a5}))
$$
r_N(\omega,x)=(\omega^2-m^2)^{(N+1)/2}
\sum\limits_{k=1}^{N+1}b_k(\omega,x)x^k,
$$
where $b_k(\omega,x)$ are uniformly bounded for $\omega\to\pm m+i0$ and any $x$.
In particular,
$$
e^{i\theta(\omega)x}=1+\sqrt{\omega^2-m^2}R_0(\omega,x)\quad\mbox{as }\,\,
 \omega\to\pm m+i0,
$$
where  $\sup\limits_{|\omega|\le m+\delta}|R_0(\omega,x)|\le C|x|$,
and  $\sup\limits_{|\omega|\le m+\delta}\Vert R_0(\omega,\cdot)\Vert_{\alpha,+}\le C<\infty$ for $\alpha<-3/2$.

%%%%%%%%%%%%%%%%%%%%%%%%%%%%%%%  Section A.2     %%%%%%%%%%%%%%%%%%
\subsection{Proof of Theorem \ref{l2.15}}\label{sec7.2}
%%%%%%%%%%%%%%%%%%%%%%%%%%%%%%%%%%%%%%%%%%%%%%%%%%%%%

%By the properties {\bf (I)}--{\bf (III)}, $K(x,t)=0$ for $t<0$.
To prove  (\ref{boundK}), we use the properties {\bf (I)}--{\bf (III)} of $\tilde K(x,\omega)$.
At first, we rewrite $K(x,t)$ in the form
 $$
 K(x,t)=\frac1{2\pi}\int\limits_{\Im\omega=\mu>0}
 e^{-i\omega t}\tilde K(x,\omega)\,d\omega
 =-\frac1{2\pi}\int\limits_{\Gamma}
 e^{-i\omega t}\tilde K(x,\omega)\,d\omega,\quad x\in\N,\quad t>0,
 $$
where $\Gamma=\{|\omega|=R:\,R>\sqrt{4+m^2}\}$, and the contour $\Gamma$ is  oriented anticlockwise.
Since $\tilde K(x,\omega)$ is analytic in $\mathbb{C}\setminus \Lambda$,
we can vary the integration contour on  $\Lambda_{\ve}$,
where  $\Lambda_{\ve}$ surrounds the segments of $\Lambda$
and belongs to an $\ve$-neighborhood of $\Lambda$
($\Lambda_{\ve}$ is  oriented anticlockwise).
Taking a limit as $\ve\to0$, we find
 \beqn\label{7.13}
 K(x,t)&=&\frac1{2\pi}\int_{\Lambda}
 e^{-i\omega t}\left(\tilde K(x,\omega+i0)-\tilde K(x,\omega-i0)\right)\,d\omega
 \nonumber\\
 &=& \frac{i}{\pi}\int_{\Lambda}
 e^{-i\omega t}\Im\tilde K(x,\omega+i0)\,d\omega
 =\sum\limits_{\pm}\sum\limits_{j=1}^2\frac{i}{\pi}\int_{\Lambda}
 e^{-i\omega t} P_j^{\pm}(x,\omega)\,d\omega.
\eeqn
Here $P_j^{\pm}(x,\omega):=\zeta_j^\pm(\omega)\Im\tilde K(x,\omega+i0)$,
 $j=1,2$, where $\zeta_j^\pm(\omega)$ are smooth functions such that
$\sum\limits_{\pm,j}\zeta_j^\pm(\omega)=1$, $\omega\in\R$,
$\supp\zeta_1^\pm\subset {\cal O}(\pm m)$,
$\supp\zeta_2^\pm\subset {\cal O}(\pm \sqrt{4+m^2})$
(${\cal O}(a)$ denotes a neighborhood of the point $\omega=a$).
In the case $m=0$, instead of $\zeta_1^\pm$
($P_1^\pm$) we introduce  the function $\zeta_1$ (respectively, $P_1$)
  with $\supp\zeta_1\subset {\cal O}(0)$.
By the property {\bf (III)},
\beqn
&&\Vert P^\pm_1(\cdot,\omega)\Vert_{\alpha,+}=O(|\omega\mp m|^{1/2})\quad\mbox{if }\,\,m\not=0,
\quad \Vert P_1(\cdot,\omega)\Vert_{\alpha,+}=O(|\omega|) \quad\mbox{if }\,\,m=0,\nonumber\\
&&\Vert P^\pm_2(\cdot,\omega)\Vert_{\alpha,+}=O(|\omega\mp\sqrt{4+m^2}|)\nonumber
\eeqn
 for any $\alpha<-3/2$. Therefore,
using  \cite[Lemma 10.2]{JK}    %%%  or  \cite[Lemma 2]{V74}
we obtain
\be\label{7.3}
\Big\Vert\int_{\Lambda}
 e^{-i\omega t} P_j^{\pm}(x,\omega)\,d\omega\Big\Vert_{\alpha,+}=O(|t|^{-3/2})
 \quad \mbox{as }\,\,t\to\infty,
 \quad\mbox{for any }\,\,\alpha<-3/2.
\ee
The bound (\ref{boundK}) follows from (\ref{7.13}) and (\ref{7.3}).\bo
%%%----------------------------------------------
%%%Formally, the proof of (\ref{7.3}) is based on the well-known estimate
%%%$\Big|\int_0^\delta e^{-i\omega t}
%%%\sqrt\omega\,d\omega\Big|\le C(1+t)^{-3/2}$
%%%(see, for example, \cite{Watson} or \cite[p.144 in Russian edition]{V74}).
%%%----------------------------------------------

%%%%%%%%%%%%%%%%%%%%%%%%%%%%%%%  Section A.3   %%%%%%%%%%%%%%%%%
\subsection{Existence of solutions}
%%-------------------------------------------------------------

\begin{lemma}\label{l2.3}
Let  $\alpha\in\R$, $m,\gamma\ge0$,  and let $P$ satisfy the condition (\ref{P}). Then
the following assertions hold.
(i) For every $Y_0\in {\cal H}_{\alpha,+}$, the problem (\ref{1.1})--(\ref{1.3})
has a unique solution $Y(t)\in C(\R,{\cal H}_{\alpha,+})$.
Moreover, the operator $U(t):Y_0\mapsto Y(t)$, $t\in \R$,
is continuous on ${\cal H}_{\alpha,+}$, and there exist constants $C,B<\infty$ such that
\be\label{2.7}
\Vert Y(t)\Vert_{\alpha,+}\le Ce^{B|t|}\quad
\mbox{for } \,\,t\in\R.
\ee
(ii) For $Y_0\in {\cal H}_{0,+}$, the following identity holds
\be\label{H-1}
{\rm H}(Y(t))+\gamma\int_0^t|\dot u(0,s)|^2\,ds={\rm H}(Y_0),\quad t\in\R,
\ee
which implies the estimate ${\rm H}(Y(t))\le{\rm H}(Y_0)$,
 where ${\rm H}(Y(t))$ is the Hamiltonian defined in (\ref{H}).
In particular, if $\gamma=0$, the energy ${\rm H}(Y(t))$ is conserved and finite.
 \end{lemma}
%%%%%%%%%%%%%%%%%%%%%%%%%%%%%%%%%%%%%%%%%%%%%%%%%%%%%%%%%%%%%
{\bf Proof of Lemma \ref{l2.3}}\,
To prove the existence of $u(x,t)$, it suffices
 to prove the existence of the solutions $q(t)\equiv q(0,t)$
 to the problem (\ref{b.2}), (\ref{b.4}).
It follows from the representation (\ref{2.1}), Lemma~\ref{l2.1}
 and formula  (\ref{qxt}).
Further, using  (\ref{qxt}), we write (\ref{b.2}) in the equivalent integral form
 \be\label{qt'}
 q(t)=\int_0^t\Big(\int_0^s{\cal F}(\tau,q(\tau))\,d\tau\Big)\,ds
 +\int_0^t\Big(\int_0^s z(1,\tau)\,d\tau-\gamma q(s)\Big)\,ds +C_0+C_1 t,\quad t>0.
 \ee
Here ${\cal F}(t,q(t)):=-(1+m^2)q(t)+F(q(t)) +\int_0^tK(1,t-s)q(s)\,ds$,
$C_0=q(0)\equiv q_0$, $C_1=\dot q(0)\equiv p_0$.
By the bound (\ref{3.7'}), condition (\ref{P}), and the contraction mapping principle,
the solution $q(t)$ of (\ref{qt'}) is unique
on a certain interval $t\in[0,\ve)$ with some  $\ve>0$ depending on the initial data $(q_0,p_0)$.
Hence,  by (\ref{qxt}),
 the solution $q(x,t)$ of the problem (\ref{b.1})--(\ref{b.4}) is unique
on a certain interval $t\in[0,\ve)$ with some $\ve>0$ depending on the initial data $Y_0$.
The existence of $z(x,t)$ is stated in Lemma \ref{l2.1}.
This implies the existence of the local solution $u(x,t)=z(x,t)+q(x,t)$ for sufficiently
small $t$. This local solution can be extended to the global solution
using the a priori  estimate (\ref{2.7}). Now we verify (\ref{2.7}).
Indeed, by (\ref{b.2}) and (\ref{qxt}),
\beqn\label{a.4}
\ba{ll}
\displaystyle
\frac12\Big(|\dot q(t)|^2+(m^2+1)|q(t)|^2\Big)+P(q(t))
+\gamma\int_0^t |\dot q(s)|^2\,ds\\
=\displaystyle \frac12\Big(|p_0|^2+(m^2+1)|q_0|^2\Big)+P(q_0)+
\int_0^t\dot q(s)\Big(z(1,s)-\int_0^s K(1,s-\tau)q(\tau)d\tau\Big)\,ds.
\ea\eeqn
 Define
$M(t)=\sup\limits_{0\le s\le t}(|q(s)|^2+|\dot q(s)|^2)$.
Then  (\ref{a.4}) and  (\ref{3.7'})  yield
$$
M(t)\le C_1+\int_0^t \sqrt{M(s)}|z(1,s)|\,ds+
C_2\int_0^t M(s)\,ds,\quad t>0.
$$
Applying the Gronwall--Bellman integral type inequality (see, for instance, \cite{Pach}),
we find
$$
M(t)\le e^{C_2 t}(\sqrt{C_1}+\frac12\int_0^t|z(1,s)|e^{C_2 s/2}\,ds)^2,\quad t>0.
$$
Since $|z(1,t)|\le C\langle t\rangle^{\sigma}\Vert Y_0\Vert_{\alpha,+}$
(see (\ref{c.2})), we obtain the a priori bound
\be\label{7.9}
|q(t)|+|\dot q(t)|\le Ce^{B|t|}
\ee
with some constants $C,B<\infty$.
 By (\ref{7.4}) and  (\ref{qxt}),
 \be\label{7.10}
\Big(\sum\limits_{x\in\N}\langle x\rangle^{2\alpha}(|q(x,t)|^2+|\dot q(x,t)|^2)\Big)^{1/2}
\le C_1e^{B|t|}.
 \ee
 Thus,  the a priori bound (\ref{2.7}) follows from (\ref{2.1}),
 (\ref{7.9}), (\ref{7.10}) and  (\ref{c.2}).
 \bo
 %%%%%%%%%%%%%%%%%%%%%%%%%%%%
%\begin{remark}\label{rem7.3}
%Let $F(q)=-\kappa q$ with $\kappa\ge0$, and $Y_0\in{\cal H}_{0,+}$.
%Then the energy ${\rm H}(Y(t))$ is nonnegative and finite,
%${\rm H}(Y(t))\le {\rm H}(Y_0)$ by (\ref{H-1}).
%\end{remark}

%%%%%%%%%%%%%%%%%%%%%%%%%    Appendix B    %%%%%%%%%%%%%%%%%%%%%%%%
\setcounter{section}{2}
\setcounter{equation}{0}
\section*{Appendix B: Properties of
$\tilde D(\omega)$ and $\tilde N(\omega)$ for $\omega\in\mathbb{C}$ }
%%%%%%%%%%%%%%%%%%%%%%%%%%%%%%%%

In this section, we study $\tilde D(\omega)$ and $\tilde N(\omega)=(\tilde D(\omega))^{-1}$.
Denote by $\mathbb{C}_+$ ($\mathbb{C}_-$) the upper (respectively, lower) half-plane,
$\mathbb{C}_\pm=\{\omega\in\mathbb{C}:\,\pm \Im\,\omega>0\}$.
%%-------------------------------------------------
\begin{lemma}\label{l2.A}
(i) The function  $\tilde N(\omega)$
is meromorphic for $\omega\in \mathbb{C}\setminus \Lambda$.

(ii) $|\tilde N(\omega)|=O(|\omega|^{-2})$ as $|\omega|\to\infty$.

(iii) $\tilde D(\omega)\not=0$  for all $\omega\in \mathbb{C}_+$.
\end{lemma}
%%-------------------------------------------------------------
{\bf Proof}\,
Assertion (i) of the lemma  follows from the formula (\ref{3.21})
and the analyticity of
$\tilde D(\omega)$  for $\omega\in \mathbb{C}\setminus \Lambda$
(see property {\bf (I)} of $\tilde K(x,\omega)$ in Appendix~A).
Assertion (ii) follows from (\ref{3.21}) and (\ref{7.0}).
To prove assertion~(iii),
we assume contrarily that $\tilde D(\omega_0)=0$ for some $\omega_0\in \mathbb{C}_+$.
Hence, the function $u_*(x,t)=e^{i\theta(\omega_0)x}e^{-i\omega_0 t}$,
$x\in\Z_+$, $t\ge0$, is a solution of the problem (\ref{1.1}), (\ref{1.2})
with the initial data $Y_*(x)=e^{i\theta(\omega_0)x}(1,-i\omega_0)$.
Therefore, the Hamiltonian (\ref{H}) is
$$
{\rm H}(u_*(\cdot,t),\dot u_*(\cdot,t))=e^{2t\,\Im\omega_0}{\rm H}(Y_*)
\quad \mbox{for all } t>0,
\quad \mbox{where }\,\, {\rm H}(Y_*)>0.
$$
Since $\Im\omega_0>0$ and $Y_*\in {\cal H}_{0,+}$,
  this exponential growth  contradicts the energy estimate (\ref{H-1}).
Hence, $\tilde D(\omega)\not=0$ for any $\omega\in \mathbb{C}_+$. \bo
%%%%%%%%%%%%%%%%%%%%%%%%%%%%%%%%%%%%%%%%%%%%%%%
\begin{cor}\label{cor-b}
If $\gamma=0$, then
$\tilde D(\omega)\not=0$ for any $\omega\in \mathbb{C}_-$.
\end{cor}

Indeed, if $\gamma=0$, then
$\overline{\tilde D(\omega)}=\tilde D(\bar\omega)$,
because $\overline{\theta(\omega)}=-\theta(\bar\omega)$ for $\omega\in \mathbb{C}\setminus \Lambda$.
Therefore, Corollary \ref{cor-b} follows from item (iii) of Lemma~\ref{l2.A}.
\medskip
%%--------------------------------------------------

Now we study the invertibility of $\tilde D(\omega)$ for $\omega\in\R$.
%%-------------------------------------
\begin{lemma}\label{l8.2}
Let $\omega\in\R$, and the constants $\gamma,m,\kappa$
satisfy condition {\bf C}.
Then $\tilde D(\omega)\not=0$ for $\omega\in\R\setminus \Lambda$ and
$\tilde D(\omega\pm i0)\not=0$ for $\omega\in\Lambda$.
\end{lemma}
%%-------------------------------------
{\bf Proof}\,
{\it Step 1}: Let $\omega\in\R$ and $|\omega|>\sqrt{4+m^2}$.
Then  $\Re\theta(\omega)=\pm\pi$.
Therefore,
$$
\tilde D(\omega)=-\omega^2+\kappa+1+m^2-i\omega\gamma+
e^{-\Im\theta(\omega)} \quad\mbox{with }\,\,\Im\theta(\omega)>0.
$$
 Hence, $\Im\tilde D(\omega)\not=0$ iff $\gamma\not=0$.
 On the other hand, $\Re\tilde D(\omega)=\kappa-2$ for $\omega=\pm\sqrt{4+m^2}$
and $\Re\tilde D(\omega_1)<\Re\tilde D(\omega_2)$ if
$|\omega_1|>|\omega_2|\ge\sqrt{4+m^2}$.
In particular, $\Re\tilde D(\omega)\to-\infty$ as $|\omega|\to\infty$.
Hence, for $|\omega|>\sqrt{4+m^2}$, $\Re\tilde D(\omega)\not=0$ iff $\kappa\le 2$.
Therefore, for such values of $\omega$,
$\tilde D(\omega)\not=0$ iff either $\gamma\not=0$ or $\gamma=0$ and $\kappa\le 2$.
If $\gamma=0$ and $\kappa>2$, then there exist two points
$\pm\omega_0$ ($\omega_0>\sqrt{4+m^2}$) such that
$\tilde D(\pm\omega_0)=0$.
\\
%%-------------------------------------
{\it Step 2}: Let $m\not=0$ and  $\omega\in(-m,m)$.
Then, $\Re\theta(\omega)=0$ and
$e^{i\theta(\omega)}\in(e^{i\theta(0)},e^{i\theta(\pm m)})=(4(m+\sqrt{4+m^2})^{-2},1)$.
%%------------------------------------------
%% $\beta_m:=2/(m+\sqrt{4+m^2})$, $m\ge0$. Note that $\beta_m=\max\limits_{\theta\in[-\pi,\pi]}|\phi'(\theta)|$,
%% where $\phi(\theta)=\sqrt{2-2\cos\theta+m^2}$. Moreover, by direct calculation, we obtain
%$\int_0^{+\infty} K(1,s)\,ds=e^{i\theta(0)}\equiv\beta_m^2$.
%%In particular, if $m\not=0$, then $\beta_m\in(0,1)$, and  $\beta_m=1$ if $m=0$.
%-----------------------------------------------------------
Hence,
$$
\Re\tilde D(\omega)=-\omega^2+\kappa+1+m^2-e^{i\theta(\omega)}
>\kappa\quad\mbox{for }\,\,|\omega|<m,
 $$
 and $\Re\tilde D(\pm m)=\kappa$.
Therefore, $\tilde D(\omega)\not=0$  for any $|\omega|<m$,
since $\kappa\ge0$.
\medskip\\
%%----------------------------------------------------
{\it Step 3}: Let $\omega\in(-\sqrt{4+m^2},-m)\cup(m,\sqrt{4+m^2})$.
Then $\Re\theta(\omega+i0)\in(-\pi,0)\cup(0,\pi)$
and $\Im\theta(\omega+i0)=0$.
Moreover, ${\rm sign}(\sin\theta(\omega+i0))={\rm sign}\,\omega$.
Hence,  for $m\not=0$,
%%$$ \Re\tilde D(\omega+i0)=-\omega^2+\kappa+1+m^2 -\cos\theta(\omega)=\kappa-\frac12(\omega^2-m^2)$$
$$
\Im\tilde D(\omega+i0)=-\omega\gamma -\sin\theta(\omega+i0)
=-{\rm sign}(\omega)\left(|\omega|\gamma+\frac12\sqrt{\omega^2-m^2}\sqrt{4+m^2-\omega^2}\right)
\not=0.
$$
If $m=0$, then for $\omega\in(-2,0)\cup(0,2)$,
$$\tilde D(\omega+i0)=\kappa-\omega^2/2-i\omega\left(\gamma+\frac12\sqrt{4-\omega^2}\right)
\not=0\quad \mbox{for any }\,\,\kappa,\gamma\ge0.
$$
%%------------------------------------
%%------------------------------------
{\it Step 4}:
Since
$\tilde D(\omega-i0)=\overline{\tilde D(\omega+i0)}-2 i\omega\gamma$
for $\omega\in\Lambda\setminus\Lambda_0$, then
\beqn
\tilde D(\omega\!-\!i0)&=&
-\omega^2+\kappa+1+m^2-
 \cos\theta(\omega+i0)+i\sin\theta(\omega+i0)-i\omega\gamma\nonumber\\
&=&
\kappa-(\omega^2-m^2)/2+i\left({\rm sign}(\omega)\frac12\sqrt{\omega^2-m^2}
\sqrt{4+m^2-\omega^2}-\omega\gamma\right),\nonumber
\eeqn
for $\omega\in\Lambda\setminus\Lambda_0$. Hence,
%%%%%%%%%%%%%%%%%%%%%%%%%%%%%%%%%%%%%%%%%%%%%%%%%%%%%%%%%%%%%5
%%It is easy to verify that
%%  there are points $\omega\in\Lambda\setminus\Lambda_0$, in which $\tilde D(\omega-i0)=0$, iff
%% $\gamma\not=0$ and one of the following conditions is fulfilled:
%% (1) $m=0$, $\kappa=2(\nu^2-\gamma^2)$ and $\gamma\in(0,\nu)$;
%% (2) $m\not=0$, $\kappa=\nu^2-\gamma^2\pm\sqrt{(\nu^2-\gamma^2)^2-m^2\gamma^2}$
%%  and $\gamma\in\left(0,\left(\sqrt{m^2+4\nu^2}-m\right)/2\right]$.
%%%%%%%%%%%%%%%%%%%%%%%%%%%%%%%%%%%%%%%%%%%%%%%%%%%%%%%%%%%%%%%%%%%%%
$\tilde D(\omega-i0)=0$ for $\omega\in\Lambda\setminus\Lambda_0$ iff
\be\label{A29}
\kappa=(\omega^2\!-\!m^2)/2\quad\mbox{and }\,\,
\sqrt{\omega^2-m^2}
\sqrt{4+m^2-\omega^2}=2|\omega|\gamma,\quad \omega^2\in(m^2,m^2+4).
\ee
Assume that $\omega$ is a solution of (\ref{A29}).
Then, evidently, $\gamma\not=0$.
Put $P:=\omega^2-m^2$. Hence, $P$ is a solution of the following equation
\be\label{A30}
P^2+4P(\gamma^2-1)+4m^2\gamma^2=0\quad\mbox{such that }\,\, P\in(0,4).
\ee
If $m=0$, then Eqn (\ref{A30}) has a unique solution $P=4(1-\gamma^2)\in(0,4)$
iff $\gamma\in(0,1)$. Then, $\kappa=(\omega^2\!-\!m^2)/2=P/2=2(1-\gamma^2)$
by the first equation in (\ref{A29}).
Thus, if $m=0$, $\kappa=2(1-\gamma^2)$ and $\gamma\in(0,1)$, then
there exist two points
$\omega=\pm\omega_*=\pm2\sqrt{1-\gamma^2}\in\Lambda\setminus\Lambda_0$
such that $\tilde D(\omega_*-i0)=0$.

If $m\not=0$, then  (\ref{A30}) has a  solution iff
$
(\gamma^2-1)^2-m^2\gamma^2\ge0$ and $\gamma\in(0,1)$.
This is equivalent to the conditions
$\gamma^2+m\gamma-1\le0$ and $\gamma\in(0,1)$, that coincides with the inequality
$\gamma\in\left(0,\left(\sqrt{m^2+4}-m\right)/2\right]$.
Therefore, if $m\not=0$ and $\gamma\in\left(0,\left(\sqrt{m^2+4}-m\right)/2\right]$, then
Eqn (\ref{A30}) has two solutions
$$
P_\pm=2(1-\gamma^2)\pm 2\sqrt{(1-\gamma^2)^2-m^2\gamma^2},\quad P_\pm\in(0, 4).
$$
Hence, $\kappa=(\omega^2\!-\!m^2)/2=P_\pm/2=1-\gamma^2\pm\sqrt{(1-\gamma^2)^2-m^2\gamma^2}$.
%%%--

Thus,
  there are points $\omega_*\in\Lambda\setminus\Lambda_0$, in which $\tilde D(\omega_*-i0)=0$, iff
 $\gamma\not=0$ and one of the following conditions is fulfilled:
 (1) $m=0$, $\kappa=2(1-\gamma^2)$ and $\gamma\in(0,1)$;
 (2) $m\not=0$, $\kappa=1-\gamma^2\pm\sqrt{(1-\gamma^2)^2-m^2\gamma^2}$
  and $\gamma\in\left(0,\left(\sqrt{m^2+4}-m\right)/2\right]$.
   These values of $\kappa,m,\gamma$ are eliminated by the condition {\bf C}.
\smallskip\\
{\it Step 5}:
If $\omega=\pm\sqrt{4+m^2}$, then $e^{i\theta(\omega)}=-1$, and
$\tilde D(\omega)=\kappa-2\mp i\gamma\sqrt{4+m^2}$.
Hence,
$$\tilde D(\pm\sqrt{4+m^2})\not=0
\quad\mbox{iff }\,\gamma\not=0
\quad \mbox{or }\,\gamma=0\quad\mbox{and }\,\kappa\not=2.
$$
If $\omega=\pm m$, then $e^{i\theta(\omega)}=1$, and
$\tilde D(\omega)=\kappa\mp i\gamma m$.
Hence, in the case $m\not=0$,
$\tilde D(\pm m)\not=0$ iff $\gamma\not=0$ or $\gamma=0$ and $\kappa\not=0$.
In the case $m=0$, $\tilde D(0)=\kappa\not=0$ iff  $\kappa\not=0$.
Therefore, if Condition {\bf C} holds, then $\tilde D(\omega_0)\not=0$ for $\omega_0\in\Lambda_0$.
\bo
%%--------------------------------------------------
\begin{cor}
If condition  {\bf C} is not satisfied, then there are points
 $\omega\in\R$, in which $\tilde D(\omega)=0$.
For example, $\tilde D(0)=0$  in the case $m=\kappa=0$.
If $\gamma=\kappa=0$, then $\tilde D(\pm m)=0$.
If $\gamma=0$ and $\kappa=2$, then $\tilde D(\pm \sqrt{m^2+4})=0$.
If $\gamma=0$ and $\kappa>2$, then $\exists\,\omega_0>\sqrt{4+m^2}$ such that
  $\tilde D(\pm \omega_0)=0$,
 and $\tilde D'(\omega_0)=-2\omega_0(\kappa-1)/(2\kappa+m^2-\omega_0^2)<0$.
\end{cor}
%%------------------------

Now we study the behavior of $\tilde D(\omega)$ and $\tilde N(\omega)=(\tilde D(\omega))^{-1}$
near the points $\omega=\pm m$ and $\omega=\pm\sqrt{4+m^2}$ with any $m,\gamma,\kappa\ge0$.
In the neighborhood of the points $\omega=\pm\sqrt{4+m^2}$ we use the representation (\ref{a9})
and obtain
\beqn\label{A.9}
\tilde D(\omega)
&=& \kappa-2\mp i\sqrt{4+m^2}\gamma- i\sqrt{4+m^2-\omega^2}
+\frac12(4+m^2-\omega^2)-i(\omega\mp\sqrt{4+m^2})\gamma
\nonumber\\
&&+\frac{i}8(4+m^2-\omega^2)^{3/2}+...,
\quad \omega\to\pm\sqrt{4+m^2},\quad \omega\in\mathbb{C}_+,
\eeqn
where ${\rm sgn}(\Re\sqrt{m^2+4-\omega^2})={\rm sgn}(\Re\omega)$.
Therefore, if $\gamma\not=0$ or $\gamma=0$ and $\kappa\not=2$, then
\be\label{B.7}
\tilde N(\omega)=(\tilde D(\omega))^{-1}=
 C_1+i\,C_2\sqrt{4+m^2-\omega^2}+\dots,
\quad\omega\to\pm\sqrt{4+m^2}, \quad\omega\in\mathbb{C}_+,
\ee
with $C_1=(\kappa-2\mp i\,\gamma\,\sqrt{4+m^2})^{-1}$ and  $C_2=C_1^2$.
If $\gamma=0$ and $\kappa=2$, then
\be\label{3.23}
\tilde N(\omega)=  i(4+m^2-\omega^2)^{-1/2}+\frac12
-\frac{i}8(4+m^2-\omega^2)^{1/2}+\dots,\quad\omega\to\pm\sqrt{4+m^2}.
\ee
%%-------------------------------------------

In the neighborhood of the points $\omega=\pm m$
we apply (\ref{a8}) (if $m\not=0$) and obtain
\be\label{B.44}
\tilde D(\omega)
= \kappa\mp i m\gamma-i\sqrt{\omega^2-m^2}-i(\omega\mp m)\gamma
-\frac12(\omega^2-m^2)+\frac{i}8(\omega^2-m^2)^{3/2}+\dots
\ee
as $\omega\to\pm m$, $\omega\in\mathbb{C}_+$,
where ${\rm sgn}(\Re\sqrt{\omega^2-m^2})={\rm sgn}(\Re\omega)$.
%%-----------------------------
In the case $m=0$,  (\ref{a10}) yields
\be\label{B.55}
\tilde D(\omega)
= \kappa-i\omega(\gamma+1)-\frac12\omega^2+\frac{i}{8}\omega^3+...,\quad \omega\to0.
\ee
%%------------------------------
Suppose that either
$m\gamma\not=0$ or $\kappa\not=0$.
Then, by (\ref{B.44}) and (\ref{B.55}), we obtain
 \be\label{B.8}
\tilde N(\omega)=(\tilde D(\omega))^{-1}=\left\{
\ba{lll}
1/\kappa+i\omega(\gamma+1)/{\kappa^2}+\dots,&\omega\to0,&{\rm if}\,\, m=0,\\
 C_3+i\,C_4\sqrt{\omega^2-m^2}+\dots,&\omega\to\pm m,&{\rm if }\,\, m\not=0,
\ea\right.\quad\omega\in\mathbb{C}_+
\ee
with $C_3=(\kappa\mp im\gamma)^{-1}$ and $C_4=C_3^2$.
If $\gamma=\kappa=0$ and $m\not=0$, then
\be\label{3.24-0}
\tilde N(\omega)= i(\omega^2-m^2)^{-1/2}-\frac12
-\frac{i}8(\omega^2-m^2)^{1/2}..., \quad \omega\to\pm m,\quad\omega\in\mathbb{C}_+.
\ee
%%---------------------------------------------------
If $m=\kappa=0$, then
\be\label{3.25}
\tilde N(\omega)= \frac{i}{\omega(\gamma+1)}-\frac{1}{2(\gamma+1)^2}
-\frac{i\omega(\gamma-1)}{8(\gamma+1)^3}+...,\quad\omega\to0.
\ee
Since $\tilde N(\omega)=(\overline{\tilde D(\bar\omega)}-2i\omega\gamma)^{-1}$
for $\omega\in\mathbb{C}_-$,
then the expansion for $\tilde N(\omega)$
as $\omega\to\omega_0$ ($\omega_0\in\Lambda_0$, $\omega\in\mathbb{C}_-$)  can be constructed
using  (\ref{A.9}) and (\ref{B.44}). In particular,
\be\label{A.11}
\tilde N(\omega+i0)-\tilde N(\omega-i0)=O(|\omega^2-\omega_0^2|^{1/2})\quad
\mbox{as }\,\,\omega\to\omega_0,\quad\omega_0\in\Lambda_0.
\ee
%%%%%%%%%%%%%%%%%%%%%%%%%%%%%%%%%%%%%%%%%%%%%%%%%%%%%%%%
{\bf Proof of Theorem \ref{l3.1}}\,
Using Lemma \ref{l2.A}, we vary the integration contour in (\ref{N}):
\be\label{b.10}
N(t)=-\frac1{2\pi}\int_{|\omega|=R}
e^{-i\omega t} \tilde N(\omega)\,d\omega,\quad t>0,
\ee
where $R$ is chosen so large  that $ \tilde N(\omega)$ has no poles
in the region $\mathbb{C}_-\cap\{|\omega|\ge R\}$.
Note that if $\gamma=0$, then $\tilde N(\omega)$ has no poles
in $\mathbb{C}_-$ by Corollary \ref{cor-b}.
Denote by $\sigma_j$ the poles of $\tilde N(\omega)$ in $\mathbb{C}_-$
(if they exist).
By Lemmas \ref{l2.A} and \ref{l8.2}, there exists a $\delta>0$ such that
$\tilde N(\omega)$ has no poles in the region $\{\omega:\Im\omega\in[-\delta,0)\}$.
Hence, we can rewrite $N(t)$ as
$$
N(t)=-i\sum\limits_{j=1}^K{\rm Res}_{\omega=\sigma_j}
\left[e^{-i\omega t} \tilde N(\omega)\right]
-\frac1{2\pi}\int_{\Lambda_{\ve}}
e^{-i\omega t} \tilde N(\omega)\,d\omega,\quad t>0,
$$
where $\varepsilon\in(0,\delta)$,
 the contour $\Lambda_{\ve}$ surrounds segments of $\Lambda$
and belongs to the $\ve$-neighborhood of $\Lambda$
($\Lambda_{\ve}$ is  oriented anticlockwise).
Passing to a limit as $\ve\to0$, we obtain
 \beqn
 N(t)&=&\frac1{2\pi}\int_{\Lambda}
 e^{-i\omega t}\left(\tilde N(\omega+i0)-\tilde N(\omega-i0)\right)\,d\omega
 +o(t^{-N})\nonumber\\
 &=&
  \sum\limits_{\pm}\sum\limits_{j=1}^2\frac{1}{2\pi}\int_{\Lambda}
 e^{-i\omega t} P_j^{\pm}(\omega)\,d\omega+o(t^{-N}),\quad t\to+\infty,
 \quad \mbox{with any }\, N>0.  \nonumber
\eeqn
Here $P_j^{\pm}(\omega):=\zeta_j^\pm(\omega)(\tilde N(\omega+i0)-\tilde N(\omega-i0))$,
where $\zeta_j^\pm(\omega)$ are as in  (\ref{7.13}).
Then (\ref{A.11})  implies the bound (\ref{NN}) with $k=0$.
Here we use the following estimate
$$
\Big|\int_{\R}\zeta(\omega) e^{-i\omega t}
(a^2-\omega^2)^{1/2}\,d\omega\Big|\le C(1+t)^{-3/2}\quad
\mbox{as }\,t\to+\infty,
$$
where $\zeta(\omega)$ is a smooth function,
and $\zeta(\omega)=1$ for $|\omega-a|\le\delta$ with some $\delta>0$
(see, for example,  \cite[Lemma 2]{V74}).
The bound (\ref{NN}) with $k=1,2$ can be proved in a similar way.
\bo
%%%%%%%%%%%%%%%%%%%%%%%%%%%%%%%%%%%%%%%%%
\begin{remark}\label{rem-b}
{\rm Now we study the asymptotics of $N(t)$ as $\to+\infty$
in the case when condition~{\bf C} is not fulfilled.
Assume first that $\gamma=\kappa=0$ and $m\not=0$.
Then $\tilde N(\omega+i0)-\tilde N(\omega-i0)=2i\Im\tilde N(\omega+i0)$,
and $\tilde N(\omega)$ has no poles in $\mathbb{C}_-$.
Introduce the circles $c_\pm$, $c_\pm=\{|\omega\mp m|=\ve\}$,
with some $\ve\in(0,m)$.
We  change the integration contour  in (\ref{b.10}) on
$\Gamma_\ve:=\cup_\pm c_\pm\cup_j\gamma_j$, where
$\gamma_j$, $j=1,2,3$, stand for the segments of the real axis
connecting the circles $c_\pm$ and passing in two directions,
$\gamma_1=[-\sqrt{m^2+4},-m-\ve]$,  $\gamma_2=[-m+\ve,m-\ve]$,
$\gamma_3=[m+\ve,\sqrt{m^2+4}]$.
Using the Cauchy theorem and Lemma \ref{l2.A}, we find
$$
 N(t)=-\frac1{2\pi}\int\limits_{c_-\cup c_+}e^{-i\omega t}\tilde N(\omega)
 \,d\omega+
\sum\limits_{j=1}^3\frac{i}{\pi}\int\limits_{\gamma_j}
e^{-i\omega t}\Im\tilde N(\omega+i0)\,d\omega.
 $$
Applying  representations (\ref{B.7}) and (\ref{3.24-0})
and  the well-known estimate (see, for example, \cite{V74})
$$
-\frac1{2\pi}\int\limits_{|\omega|=m+1}e^{-i\omega t}
(\omega^2-m^2)^{-1/2}\,d\omega=i\,\sqrt{\frac{2}{\pi m t}}\,\cos(mt-\pi/4)
+O(t^{-3/2}),\quad t\to\infty,
$$
 we obtain that
 $N(t)=-(2/(\pi m))^{1/2} t^{-1/2}\cos(mt-\pi/4)+O(t^{-3/2})$ as $t\to\infty$.

Similarly, if $\gamma=0$ and $\kappa=2$, then (\ref{3.23}) and (\ref{B.8})
give the bound $|N(t)|\le C\langle t\rangle^{-1/2}$.
 %$N(t)=(2/(\pi \sqrt{m^2+4}))^{1/2} t^{-1/2}\cos(t\sqrt{m^2+4}-\pi/4)+O(t^{-3/2})$
% as $t\to\infty$.

In the case  $\gamma=0$ and $\kappa>2$, $\tilde N(\omega)$ has two simple poles in the points
$\pm\omega_0$ with $\omega_0>\sqrt{4+m^2}$.
Then, calculating the rescue of the function $e^{-i\omega t}\tilde N(\omega)$ in
these points, we have $N(t)\sim C_1\sin\omega_0 t+O(t^{-3/2})$ as $t\to+\infty$
with some constant $C_1\not=0$.

If $m=\kappa=0$, then we use formulas (\ref{3.25}) and (\ref{B.7}),
  calculate the rescue of $\tilde N(\omega)$ in the point
$\omega=0$ and obtain $N(t)=(\gamma+1)^{-1}+O(t^{-3/2})$ as $t\to+\infty$.

Finally, let one of the following two conditions hold (see condition {\bf C}):
\begin{itemize}
  \item[(1)] $m=0$, $\kappa=2(1-\gamma^2)$ and $\gamma\in(0,1)$
  \item[(2)]
  $m\not=0$, $\kappa=1-\gamma^2\pm\sqrt{(1-\gamma^2)^2-m^2\gamma^2}$
   and $\gamma\in\left(0,\left(\sqrt{m^2+4}-m\right)/2\right]$
 \end{itemize}
  Then there exist points $\omega_*\in\Lambda\setminus\Lambda_0$ such that
  $\tilde D(\omega_*-i0)=0$ (see Step~4 in the proof of Lemma~\ref{l8.2}).
%%We denote  $\theta_+:=\lim_{\ve\to+0}\theta(\omega_*+i\ve)$, $\theta_+\in\R$.
%%  Then the function of the form $u(x,t)=\sin(\theta_+x+\omega_* t)$
%%  is a solution of the system (\ref{1.1})--(\ref{1.2}).
Hence,  $N(t)=C\sin(\omega_* t)+O(t^{-3/2})$ as $t\to\infty$ with some constant $C\not=0$.
%If $Y_0(0)\not=0$, then by (\ref{3.15}), we have
% $u(0,t)=q(0,t)=C\sin(\omega_* t+\beta)+O(t^{-3/2})$ as $t\to\infty$,
% where $C_1\not=0$.
}
\end{remark}

%%%%%%%%%%%%%%%%%%%%%%%%%%    Appendix C %%%%%%%%%%%%%%%%%%%%%%%%
\setcounter{section}{3}
\setcounter{equation}{0}
\setcounter{theorem}{0}
\section*{Appendix C: Zero boundary condition }
%%%%%%%%%%%%%%%%%%%%%%%%%%%%%%%%

Consider the following initial-boundary value problem on the half-line:
\beqn\label{B.2}
\left\{\ba{l}
\ddot u(x,t)=(\Delta_L-m^2)u(x,t),\quad x\in\N,\quad t\in\R,\\
u(0,t)=0, \\
u(x,0)=u_0(x),\quad \dot u(x,0)=v_0(x),\quad x\in\N.
\ea\right.
\eeqn
In \cite{D08}, we have proved the convergence to equilibrium
for the harmonic crystals on the half-space in any dimension with zero boundary condition.
However, the one-dimensional case with $m=0$ was not considered.
Therefore, we outline the strategy of the proof of Theorem~\ref{t1}.

At first, we rewrite the problem (\ref{B.2}) in the  more general form:
%%---------------------------------------
\beqn\label{C.1}
\left\{\ba{l}
\ddot u(x,t)=-\sum\limits_{x'\ge0}\left(V(x-x')-V(x+x')\right)
u(x',t),\quad x\in\N,\quad t\in\R,\\
u(0,t)=0, \\
u(x,0)=u_0(x),\quad \dot u(x,0)=v_0(x),\quad x\in\N.
\ea\right.
\eeqn
%%--------------------
We assume that $u_0(0)=v_0(0)=0$.
We impose the following conditions {\bf V1}--{\bf V5} on the interaction function $V$.

\begin{description}
\item{\bf V1} There exist positive constants $C$ and $\beta$ such that
$|V(x)|\le C e^{-\beta|x|}$ for $x\in \Z$.

\item{\bf V2} $V(x)$ is real and even, i.e., $V(-x)=V(x)\in \R$, $x\in \Z$.
\end{description}

Conditions {\bf V1} and {\bf V2} imply that $\hat V(\theta)$ is a real-analytic
 function of $\theta\in \T\!$.
\begin{description}
\item{\bf V3}  $\hat V(\theta)\ge0$ for every $\theta \in \T$.
\end{description}

Let us define the real-valued nonnegative function,
$\phi(\theta)=\big(\hat V(\theta )\big)^{1/2}\ge 0$.
 $\phi(\theta)$ can be chosen as the real-analytic function
in $\T\setminus{\cal C}_*$,
where %${\cal C}_*$ is  a closed subset of $\T$ such that
the Lebesgue measure of a set ${\cal C}_*$ is zero
 (see Lemma 2.2 in \cite{DKS1}).

\begin{description}
 \item{\bf V4} $\phi''(\theta)$ does not vanish  identically
on $\theta \in\T\setminus {\cal C}_*$.

  \item{\bf V5}
  $\hat V(\theta)\not=0$ for all $\theta\not=0$,
and  $\hat V(\theta)\ge C\theta^2$ as $\theta\to0$ with some positive constant $C$.
\end{description}

Conditions {\bf V1}--{\bf V5}  are fulfilled, for example,
in the case when  $V(x)$ has a form
\be\label{V}
V(\pm1)=-1,\quad V(x)=0\,\,\, \mbox{for }\, |x|\ge2,\quad
\mbox{and }\,\,
\quad V(0)=2+m^2\,\,\,\mbox{with }\,\,m\ge0.
\ee
 Then $\phi(\theta)=\sqrt{2-2\cos\theta+m^2}$.
If $m\not=0$, then ${\cal C}_*=\emptyset$.        %%% and $\hat V(\theta)\not=0$ for all $\theta\in\T$
If $m=0$, then $\phi(\theta)=2|\sin(\theta/2)|$ and ${\cal C}_*=\{0\}$.  %%% and $\hat V(0)=0$.
Furthermore, in the case when $V(x)$ is of the form (\ref{V}),
 the problem (\ref{C.1}) becomes (\ref{B.2}).
\medskip

Write $Z(x,t)=\left(Z^0(x,t),Z^1(x,t)\right)\equiv(u(x,t),\dot u(x,t))$,
$Z_0(x)=(u_0(x),v_0(x))$.
The existence of dynamics is stated by the following lemma which is proved as
 in \cite{D08}.
 %%%%%%%%%%%%%%%%%%%%%%%%%%%%%%%%%%%
\begin{lemma}
Let $\alpha\in\R$ and conditions {\bf V1} and {\bf V2} hold.
For any $Z_0\in{\cal H}_{\alpha,+}$, there exists a unique solution
$Z(\cdot,t)\in {\cal H}_{\alpha,+}$ to problem (\ref{C.1}).
Moreover, the operator $U_0(t):Z_0\to Z(\cdot,t)$
 is continuous in ${\cal H}_{\alpha,+}$.
\end{lemma}
%%---------------------------------

Indeed, by condition {\bf V2}, the solution of  problem (\ref{C.1})
can be represented as the restriction of the solution to the Cauchy problem
with odd initial data on the half-line,
$$
Z^i(x,t)=\sum\limits_{y\in\Z}
{\cal G}^{ij}_{t}(x-y) Z^j_{\rm odd}(y),\quad
x\in\Z_+,\quad i=0,1.
$$
Here ${\cal G}_t(x)=F^{-1}_{\theta\to x}[e^{\hat {\cal A}(\theta)t}]$
with $\hat{\cal A}(\theta)=\left(\ba{ccc}0&1\\\hat V(\theta)&0\ea\right)$
 (see also formulas (\ref{3.2}), (\ref{hatcalG})
 with $\phi(\theta)=(\hat V(\theta))^{1/2})$,
 and by definition,
$Z_{\rm odd}(x)=Z_0(x)$ for $x>0$, $Z_{\rm odd}(0)=0$,
and $Z_{\rm odd}(x)=-Z_0(-x)$ for $x<0$.
Then the solution $Z(x,t)$ of  problem (\ref{C.1})
is of a form
\be\label{10.12}
Z^i(x,t)=\sum\limits_{y\in\N}
\Big({\cal G}^{ij}_t(x-y)-{\cal G}^{ij}_t(x+y)\Big)Z^j_0(y),\quad
x\in\Z_+,\quad i=0,1.
\ee
Furthermore, the following bound holds,
\be\label{c.2}
\Vert U_0(t) Z_0\Vert_{\alpha,+}\le C\langle t\rangle^\sigma\Vert Z_0\Vert_{\alpha,+},
\ee
with some constants $C=C(\alpha),\sigma=\sigma(\alpha)<\infty$.

Denote by $\nu_0$ a Borel probability measure on ${\cal H}_{\alpha,+}$
giving the distribution of $Z_0$.
We impose conditions {\bf S1}--{\bf S4} on $\nu_0$.
 Let $\nu_t$, $t\in\R$, denote the distribution of
the solution $Z(x,t)=U_0(t)Z_0$ (see Definition~\ref{def2.6}).
Then the following theorem holds (cf Theorem~\ref{t1}).
%%%%%%%%%%%%%%%%%%%%%%%%%%%%%%%%%%%%%%%%%%%%%%%%%%%%%%%%%%%%%%%%%%%%%
\begin{theorem}\label{t9.2}
Let conditions {\bf S1}--{\bf S4} and {\bf V1}--{\bf V5} hold,
and let $\alpha<-1/2$ if $\hat V(0)\not=0$,
and $\alpha<-1$ if $\hat V(0)=0$.
 Then the following assertions are true.\\
(i)  The measures $\nu_t$ weakly converge
on the space ${\cal H}_{\alpha,+}$ as $t\to\infty$.
Moreover, the limit measure $\nu_\infty$ is Gaussian on ${\cal H}_{\alpha,+}$
with the correlation matrix of the form (\ref{correlation})--(\ref{q-infty})
with $\phi(\theta)=(\hat V(\theta))^{1/2}$.\\
(ii) The correlation functions of $\nu_t$
converge to a limit as $t\to\infty$, i.e., (\ref{concor}) holds.
\end{theorem}
%%%----------------------------------------------------

The derivation of this theorem is based on the proof of the
compactness of the measures family $\{\nu_t,t\in\R\}$ and the convergence
of the characteristic functionals.
Below we prove only the compactness.  The convergence of
the characteristic functionals and  correlation functions can be proved
using the technique from \cite{D08}.
 The compactness  follows from the bound (\ref{20.1}) below
by the Prokhorov compactness theorem \cite[Lemma II.3.1]{VF} and
by a method used in \cite[Theorem XII.5.2]{VF}.
%% since the embedding  ${\cal H}_{\alpha,+}\subset {\cal H}_{\beta,+}$ is compact if $\alpha>\beta$.
%%%%%%%%%%%%%%%%%%%%%%%%%%%%%%%%%%%%%%%%%%%%%%%%%%%%
\begin{lemma}
Let all assumptions of Theorem \ref{t9.2} be fulfilled.
Then
 \be \label{20.1}
\sup\limits_{t\ge 0}\E_0\left(\Vert U_0(t)Z_0\Vert^2_{\alpha,+}\right)<\infty.
\ee
\end{lemma}
%%%%%%%%%%%%%%%%%%%%%%%%%%%%%%%%%%%%%%%%%%%%%%%%%%%%%%%%%%%%%%%%
{\bf Proof}. By  Definition \ref{d1.1'},
\be\label{10.7}
\E_0 \left(\Vert  Z(\cdot,t)\Vert^2_{\alpha,+}\right)=
\!\sum\limits_{z\in\Z_+} (1+|z|^2)^\alpha
\Big(Q_t^{00}(z,z)+Q_t^{11}(z,z)\Big).
\ee
The representation (\ref{10.12}) gives
$$
Q^{ij}_t(z,z')=\E_0\Big(Z^i(z,t)\otimes Z^j(z',t)\Big)
= \langle Q_0(y,y'), {\bf G}^i_{z}(y,t)\otimes
{\bf G}^j_{z'}(y',t)\rangle_+,
$$
where ${\bf G}^i_{z}(y,t)$ is defined in (\ref{2.26}).
If $\hat V(0)\not=0$, then the Parseval identity
and formula (\ref{hatcalG})
 with $\phi(\theta)=(\hat V(\theta))^{1/2}$ yield
\beqn\label{10.8}
\Vert{\bf G}^i_{z}(\cdot,t)\Vert^2_{l^2}
%%%= \frac1{2\pi}\int_{\T} |\hat{\bf G}^i_{z}(\theta,t)|^2\,d\theta
\le C\int_{\T}\Big( |\hat{\cal G}^{i0}_t(\theta)|^2
+|\hat{\cal G}^{i1}_t(\theta)|^2\Big)\sin^2(z\theta)\,d\theta \le C_0<\infty.
\eeqn
By conditions {\bf S1}, {\bf S2} and {\bf S4},
the bound (\ref{2.42}) holds. Therefore,
\beqn\label{10.6}
\ba{rcl}
|Q^{ij}_t(z,z')|&=&\left|\langle Q_0(y,y'), {\bf G}^i_{z}(y,t)\otimes
{\bf G}^j_{z'}(y',t)\rangle_{+}\right|\\
&\le& C\Vert{\bf G}^i_{z}(\cdot,t)\Vert_{0,+}\,
\Vert{\bf G}^j_{z'}(\cdot,t)\Vert_{0,+}\le C_1<\infty,
\ea\eeqn
where the constant $ C_1$  does  not depend on
$z,z'\in\Z_+$ and $t\in\R$.
Therefore, (\ref{20.1}) follows from (\ref{10.7}) and (\ref{10.6}),
since $\alpha<-1/2$.
\smallskip

If  $\hat V(0)=0$, then $\hat V(\theta)\ge C\theta^2$ as $\theta\to0$.
The estimate (\ref{10.8}) with $i=1$ remains true.
Then $|Q_t^{11}(z,z')|\le C<\infty$, by (\ref{10.6}).
However, since
$\hat {\cal G}_t^{01}(\theta)=\sin(\phi(\theta)t)/\phi(\theta)$, we have
$$
\Vert{\bf G}^0_{z}(\cdot,t)\Vert^2_{l^2}\le
C+C_1\int_{\T}\frac{\sin^2(\phi(\theta)t)}{\hat V(\theta)}\sin^2(z\theta)\,d\theta
\le C+C_2\int_{{\cal O}(0)}\frac{\sin^2(z\theta)}{\sin^2(\theta)}\,d\theta
\le C+C_3|z|,
$$
uniformly on $t\in\R$.
Hence,
$|Q^{00}_t(z,z)|\le C\Vert{\bf G}^0_z(\cdot,t)\Vert^2_{0,+}\le C_1+C_2|z|$. %by (\ref{10.6}).
Therefore, for any $\alpha<-1$, we have
$\E_0 \left(\Vert  Z(\cdot,t)\Vert^2_{\alpha,_+}\right)\le
\sum\limits_{z\in\Z_+} \langle z\rangle^{2\alpha}(C_1+C_2|z|)<\infty$.
\bo

%%%%%%%%%%%%%%%%%%%%%%%%%%%%%%%%%%%%%%

\end{document}